\documentclass[11pt, onecolumn]{IEEEtran}
\usepackage{soul}
\usepackage[mathscr]{eucal}
\usepackage[cmex10]{amsmath}
\usepackage{epsfig,epsf,psfrag}
\usepackage{amssymb,amsmath,amsthm,amsfonts,latexsym}
\usepackage{amsmath,graphicx,bm,xcolor,url,overpic}
\usepackage{fixltx2e}
\usepackage{array}
\usepackage{verbatim}
\usepackage{bm}
\usepackage{algorithmic}
\usepackage{algorithm}
\usepackage{verbatim}
\usepackage{textcomp}
\usepackage{mathrsfs}
\usepackage{epstopdf}

\newcommand{\openone}{\leavevmode\hbox{\small1\normalsize\kern-.33em1}}

\catcode`~=11 \def\UrlSpecials{\do\~{\kern -.15em\lower .7ex\hbox{~}\kern .04em}} \catcode`~=13

\allowdisplaybreaks[4]

\newcommand{\nn}{\nonumber}

\newcommand{\calB}{\mathcal{B}}
\newcommand{\calC}{\mathcal{C}}

\newcommand{\calF}{\mathcal{F}}

\newcommand{\calP}{\mathcal{P}}

\newcommand{\calR}{\mathcal{R}}

\newcommand{\calT}{\mathcal{T}}
\newcommand{\calU}{\mathcal{U}}

\newcommand{\calX}{\mathcal{X}}
\newcommand{\calY}{\mathcal{Y}}
\newcommand{\calZ}{\mathcal{Z}}

\newcommand{\bF}{\mathbf{F}}
\newcommand{\bg}{\mathbf{g}}

\newcommand{\bx}{\mathbf{x}}
\newcommand{\bX}{\mathbf{X}}
\newcommand{\by}{\mathbf{y}}
\newcommand{\bY}{\mathbf{Y}}

\newcommand{\bZ}{\mathbf{Z}}

\newcommand{\rmA}{\mathrm{A}}

\newcommand{\rmB}{\mathrm{B}}
\newcommand{\rmc}{\mathrm{c}}
\newcommand{\rmC}{\mathrm{C}}

\newcommand{\rme}{\mathrm{e}}

\newcommand{\rmU}{\mathrm{U}}

\newcommand{\bbE}{\mathbb{E}}
\newcommand{\bbI}{\openone}

\newcommand{\bbN}{\mathbb{N}}

\newcommand{\bbR}{\mathbb{R}}

\DeclareMathAlphabet{\mathbsf}{OT1}{cmss}{bx}{n}
\DeclareMathAlphabet{\mathssf}{OT1}{cmss}{m}{sl}

\DeclareSymbolFont{bsfletters}{OT1}{cmss}{bx}{n}
\DeclareSymbolFont{ssfletters}{OT1}{cmss}{m}{n}
\DeclareMathSymbol{\bsfGamma}{0}{bsfletters}{'000}
\DeclareMathSymbol{\ssfGamma}{0}{ssfletters}{'000}
\DeclareMathSymbol{\bsfDelta}{0}{bsfletters}{'001}
\DeclareMathSymbol{\ssfDelta}{0}{ssfletters}{'001}
\DeclareMathSymbol{\bsfTheta}{0}{bsfletters}{'002}
\DeclareMathSymbol{\ssfTheta}{0}{ssfletters}{'002}
\DeclareMathSymbol{\bsfLambda}{0}{bsfletters}{'003}
\DeclareMathSymbol{\ssfLambda}{0}{ssfletters}{'003}
\DeclareMathSymbol{\bsfXi}{0}{bsfletters}{'004}
\DeclareMathSymbol{\ssfXi}{0}{ssfletters}{'004}
\DeclareMathSymbol{\bsfPi}{0}{bsfletters}{'005}
\DeclareMathSymbol{\ssfPi}{0}{ssfletters}{'005}
\DeclareMathSymbol{\bsfSigma}{0}{bsfletters}{'006}
\DeclareMathSymbol{\ssfSigma}{0}{ssfletters}{'006}
\DeclareMathSymbol{\bsfUpsilon}{0}{bsfletters}{'007}
\DeclareMathSymbol{\ssfUpsilon}{0}{ssfletters}{'007}
\DeclareMathSymbol{\bsfPhi}{0}{bsfletters}{'010}
\DeclareMathSymbol{\ssfPhi}{0}{ssfletters}{'010}
\DeclareMathSymbol{\bsfPsi}{0}{bsfletters}{'011}
\DeclareMathSymbol{\ssfPsi}{0}{ssfletters}{'011}
\DeclareMathSymbol{\bsfOmega}{0}{bsfletters}{'012}
\DeclareMathSymbol{\ssfOmega}{0}{ssfletters}{'012}

\newcommand{\hatK}{\hat{K}}

\newcommand{\hatP}{\hat{P}}

\newcommand{\tilP}{\tilde{P}}

\newcommand{\hatQ}{\hat{Q}}

\newcommand{\hatW}{\hat{W}}

\newcommand{\barX}{\bar{X}}
\newcommand{\barY}{\bar{Y}}
\newcommand{\barZ}{\bar{Z}}

\DeclareMathOperator{\var}{\mathsf{Var}}

\newtheorem{theorem}{Theorem}
\newtheorem{lemma}[theorem]{Lemma}

\newtheorem{corollary}[theorem]{Corollary}
\newtheorem{definition}{Definition}
\newtheorem{example}{Example}

\newtheorem{remark}{Remark}

\newcommand{\E}[2][]{{\mathbb{E}_{#1}}{\left[#2\right]}}       
\renewcommand{\P}[2][]{{\Pr}_{#1}{\left\{#2\right\}}}
\newcommand{\Var}[1]{{\text{\textnormal{Var}}{\left(#1\right)}}}       
\newcommand{\D}[2]{{{D}\!\left({#1\Vert#2}\right)}}
\newcommand{\V}[2]{{{V}\!\left({#1,#2}\right)}}

\newcommand{\Exp}[1]{\ensuremath{\exp\left(#1\right)}}
\newcommand{\abs}[1]{\ensuremath{\left|#1\right|}}              

\usepackage{stfloats}
\usepackage{url}
\usepackage{bbm}
\usepackage{verbatim}
\usepackage{cite}
\usepackage{enumerate}
\usepackage[ colorlinks = true,
linkcolor = blue,
urlcolor  = blue, 
citecolor = red,
anchorcolor = green,]{hyperref}

\title{Capacity Region for Covert Secret Key Generation
over Multiple Access Channels}

\author{Yingxin Zhang, Lin Zhou and Qiaosheng Zhang

\thanks{Yingxin Zhang and Lin Zhou are with the School of Cyber Science and Technology, Beihang University, Beijing, China, 100191 (Emails: \{zhang\_yx, lzhou\}@buaa.edu.cn).}
\thanks{Qiaosheng Zhang is with the Shanghai AI Laboratory (Email: zhangqiaosheng@pjlab.org.cn).}
}

\begin{document}
\maketitle
\begin{abstract}
We study covert secret key generation over a binary-input two-user multiple access channel with one-way public discussion and derive bounds on the capacity region. Specifically, in this problem, there are three legitimate parties: Alice, Bob and Charlie. The goal is to allow Charlie to generate a secret key with Alice and another secret key with Bob, reliably, secretly and covertly. Reliability ensures that the key generated by Alice and Charlie is the same and the key generated by Bob and Charlie is the same. Secrecy ensures that the secret keys generated are only known to specific legitimate parties. Covertness ensures that the key generation process is undetectable by a warden Willie. As a corollary of our result, we establish bounds on the capacity region of wiretap secret key generation without the covertness constraint and discuss the impact of covertness. Our results generalize the point-to-point result of Tahmasbi and Bloch (TIFS 2020) to the setting of multiterminal communication.
\end{abstract}


\section{Introduction}
Secret key generation ~\cite{Ahlswede1993KG, Maurer1993KG} is a longstanding area of research, where two legitimate parties, Alice and Bob, aim to generate a key using correlated source sequences so that the eavesdropper Eve cannot obtain the key. Two models for secret key generation include the \emph{source model} and the \emph{channel model}. In the source model \cite{GohariKG1}, Alice, Bob, and Eve access correlated source sequences $(X^n,Y^n,Z^n)$, respectively. In the channel model \cite{Gohari2010Channel}, there is a discrete memoryless channel $Q_{YZ|X}$, where Alice controls the channel input, while Bob and Eve observe the channel outputs at their respective ends. In both models, after obtaining correlated samples, Alice and Bob communicate interactively over an authenticated noiseless public channel to generate the secret key. For a more detailed discussion, the reader can refer to the classical textbook by Bloch and Barros~\cite{Bloch2011PhysicalLayer}.

As the communication systems continuously evolve, in 5G and beyond, multiuser communications become increasingly important, which necessitates the need for multiterminal secure communication. Towards this goal, Csisz\'ar and Narayan~\cite{Csiszar2004Source,Csiszar2008Channel,Csiszar2014MAC} initiated the study of multiterminal key generation, enabling multiple terminals to generate a common secret key simultaneously. Three problems were identified based on the secrecy constraints of the key: (i) \emph{secret key (SK) generation}, where the key is concealed only from public messages transmitted by legitimate users during interactive communication, (ii) \emph{private key (PK) generation}, where the key is concealed from both the public messages and observed source sequences of untrusted helpers, and (iii) \emph{wiretap secret key (WSK) generation}, where the key is hidden from both the public messages and the observed source sequence of the eavesdropper. The capacity region for SK and {PK} generation have been characterized completely, while only inner and outer bounds have been derived for WSK generation. Other studies on the multiple key generation problem include \cite{Jafari2017,Xu2016,Tu2017}.

Models for generating multiple keys simultaneously have also been studied. Specifically, for the source model, Ye~\cite{Ye2005InformationTG} characterized the capacity region {for PK and secret-private key generation}, which was later refined by Zhang~\emph{et al.}~\cite{Huishuai2014Three, Huishuai2017Cellular} and generalized to a cellular model. Subsequently, Zhou~\cite{Zhou2020} generalized the results in~\cite{Huishuai2014Three} to the continuous case, where each observed sample is a continuous sequence generated from an arbitrary distribution. In contrast, corresponding results for channel models are relatively few and incomplete. Salimi~\emph{et al.}~\cite{Salimi2011MAC} studied the problem of two-terminal secret key generation over a multiple access channel (MAC), while Gohari and Kramer~\cite{Gohari2023Upper} derived an outer bound on the capacity region for multiterminal WSK generation. As a corollary of our result, we establish an inner bound to the problem in~\cite{Gohari2023Upper}.

Although the above studies ensure that the generated keys are unavailable to the eavesdropper, such a guarantee is not sufficient in certain sensitive communication scenarios, where the key generation process should remain undetected (e.g., communication between a submarine and command center). To solve this problem, based on WSK generation, Tahmasbi and Bloch~\cite{Tahmasbi2017cns} initiated the study of \emph{covert secret key (CSK) generation} over a channel model. In addition to the reliability and secrecy constraints in WSK generation, an additional covertness constraint was introduced to ensure that the key generation process is undetected by the warden Willie. Bounds on the CSK capacity were derived~\cite{Tahmasbi2017cns,Tahmasbi2020KGActive}. The key idea is to combine the analysis of WSK generation and covert communication~\cite{Che2013, Wang2016, Bloch2016Resolv}. Since the covertness constraint is introduced, the key rate is no longer positive and scales in the order of reciprocal of the square root of the sequence length.

Despite its importance, multiterminal CSK generation has \emph{not} been studied. To fill the research gap, we study CSK generation over a binary-input MAC and derive bounds on the key capacity region. Specifically, in our problem, two legitimate parties, Alice and Bob, covertly generate two secret keys with the third legitimate party Charlie. The key generation process should be undetected by the warden Willie. Our CSK problem finds application in scenarios where multiple military subordinates aim to generate secret keys with the command center at the same time, while ensuring that the key generation process is undetectable by any malicious party. Subsequently, the generated keys enable covert communication between the subordinates and the command center. Our main contributions are summarized in the next subsection.

\subsection{Main Contributions}
To support multiuser covert communication, we study the problem of covert secret key generation over a binary-input multiple access channel with one-way public discussion and derive bounds on the key capacity region. When the covertness constraint is imposed, the key rate is no longer positive. Instead, the number of keys generated using source sequences of length $n$ scales in the order of $O(\sqrt{n})$, leading to key rates scaling as $O(\frac{1}{\sqrt{n}})$. Our main results concern upper and lower bounds on the pre-constants of the rates of the two generated keys. Two numerical examples are provided to illustrate our results. When the covertness constraint is removed, our results specialize to capacity region for multiterminal WSK generation. Comparing the results with and without the covertness constraint, we discuss the impact of covertness on the capacity region of multiterminal key generation.

To derive the achievability result, we adopt the \emph{likelihood encoder technique} in \cite{Song2016LikelihoodEn} and adapt the point-to-point framework in \cite{Tahmasbi2017cns} to design a coding scheme for an auxiliary problem (cf. Lemma \ref{lemma:auxi}). {We remark that the generalization is nontrivial since the communication direction for key generation to accommodate multiple users is inverted compared to point-to-point case (cf. Remark 1 on Page 3 for details).}
The auxiliary problem is connected to the original problem and enables us to decompose the performance analysis into five parts: source simulation, reliability, secrecy, covertness, and key rate. The first three parts are analyzed by judiciously applying non-asymptotic results for channel coding and channel resolvability \cite[Appendices D and E]{Arumugam2019MAC} over a MAC. The remaining two parts are analyzed by carefully designing the input distribution via the covert process in~\cite[Section IV]{Arumugam2019MAC}. As shown in our achievability proof, the key rates in our problem correspond to the secret key rates required for covert communication over the same MAC, and both rates correspond to the difference between the rates required to achieve channel reliability and channel resolvability. To prove the converse part, we apply the results for multiterminal key generation by Csisz\'ar and Narayan \cite{Csiszar2004Source,Csiszar2008Channel}, with appropriate modifications to deal with the covertness constraint.

\subsection{Organization of the Paper}
{The rest of the paper is organized as follows. In Section \ref{section:ProblemFormulation}, we set up the notation and formulate the CSK generation problem. In Section \ref{section:MainResult}, we present and discuss main results. The proofs of our results are provided in Sections \ref{section:mainproof_ach} and \ref{section:mainproof_con}. Finally, in Section \ref{section:conclusion}, we conclude our paper and discuss future research directions. For smooth presentation of our main results, the proofs of all supporting lemmas are deferred to the appendices.}

\section{Problem Formulation}
\label{section:ProblemFormulation}
\subsection*{Notation}
{Random variables (RV) are denoted by upper case letters (e.g., $X$), while their realizations are denoted by lowercase (e.g, $x$). Vectors are denoted by boldface fonts (e.g., $\bX$ and $\bx$). All sets are denoted in calligraphic font (e.g., $\calX$). Given any set $\calX$, we use $\calX^\rmc$ to denote its complement. We use $\log$ and $\exp$ with base $2$. We use $\bbR$, $\bbR_{+}$ and $\bbN$, $\bbN_+$ to denote the sets of real numbers, positive real numbers, natural numbers, and positive natural numbers, respectively. For real number $x\in\bbR$, we use $\left\{x\right\}^{+}$ to denote $\max\left\{0,x\right\}$.  Given any two integers $(a,b)\in\bbN^2$ such that $a\leq b$, we use $[a:b]$ to denote the set of all integers between $a$ and $b$, and use $[a]$ to denote $[1:a]$ for any integer $a\geq 1$. Given any $T\in\bbN_{+}$, let $\calT:=[T]$, and we use $X_\calT$ to denote the collection of RVs $X_1,\cdots,X_T$. Let $\calP(\calX)$ denote the set of all distributions over the alphabet $\calX$ and let $\calP(\calY|\calX)$ denote the set of all conditional distributions from $\calX$ to $\calY$. Given any two distributions $(P,Q)\in\calP(\calX)^2$, we use $P\ll Q$ to mean that $P$ is absolutely continuous with respect to $Q$, i.e., for all $x\in\calX$, $P(x)=0$ if $Q(x)=0$. Furthermore, we use $\D{P}{Q}:=\sum_x P(x)\log\frac{P(x)}{Q(x)}$ to denote the KL divergence, $\V{P}{Q}:=\frac{1}{2}\sum_x\left|P(x)-Q(x)\right|$ to denote the TV distance and use $\chi^2(P||Q):=\sum_x \frac{(P(x)-Q(x))^2}{Q(x)}$ to denote the Chi-squared divergence, respectively.
We use $P^{\rmU}_X$ to denote the uniform distribution over $\calX$. The binary entropy function is defined as $H_b(x):=-x\log x-(1-x)\log(1-x)$.} {Finally, we follow \cite[Section 3.1]{2009Algorithms} for the asymptotic notation including $O(\cdot)$, $\Theta(\cdot)$, $\omega(\cdot)$, follow \cite{CoverElements} for information theoretical quantities, and follow \cite{Boucheron2013ineqbook} for concentration inequalities.}

\begin{figure}[tb]
\centering
\includegraphics[width=0.6\textwidth]{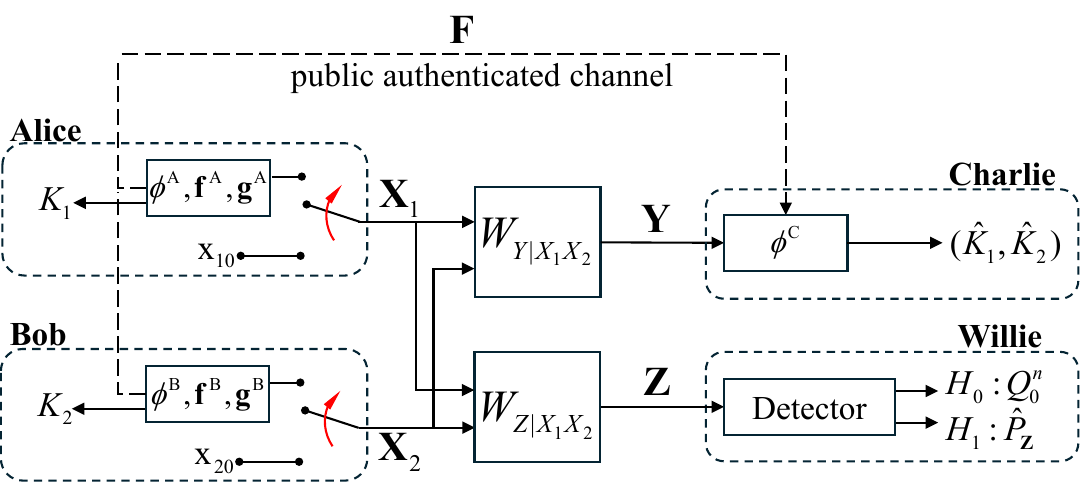}
\caption{System model for covert secret key generation over a two-user multiple access channel.}
\label{fig:orig_model}
\end{figure}

\subsection{Covert Secret Key Generation}
The problem of CSK generation over a MAC is illustrated in Fig.~\ref{fig:orig_model}. Specifically, two legitimate users Alice and Bob aim to generate secret keys $K_1,K_2$, respectively, with another legitimate user Charlie, while the warden Willie aims to detect whether the key generation process is running or not. Consistent with \cite[Section 4]{Bloch2011PhysicalLayer}, we allow the legitimate parties to randomize the transmitted messages via local randomness. Furthermore, public discussion over a noiseless channel is also allowed for legitimate users, enabling them to generate the secret keys by exchanging necessary information.

{Fix integers $(n,M_1,M_2)\in\bbN_+^3$ and finite sets $(\calX_1,\calX_2,\calY,\calZ,\calF)$. The legitimate parties Alice, Bob and Charlie communicate through a discrete memoryless MAC $W_{Y|X_1X_2}\in\calP(\calY|\calX_1\calX_2)$ to generate correlated sequences $\bX_1,\bX_2,\bY$, each of length $n$. The warden Willie observes a correlated sequence $\bZ$ of length $n$ via another MAC $W_{Z|X_1X_2}\in\calP(\calZ|\calX_1\calX_2)$. 
To facilitate key generation, local randomness is provided at all legitimate users: Alice has local randomness $R_\rmA$ generated from a distribution $P_{R_{\rmA}}\in\calP(\calR_\rmA)$, Bob  has local randomness $R_\rmB$ generated from a distribution $P_{R_{\rmB}}\in\calP(\calR_\rmB)$, Charlie has local randomness $R_\rmC$ generated from a distribution $P_{R_{\rmC}}\in\calP(\calR_\rmC)$. The following definition specifies how keys are generated.
\begin{definition}\label{def:KGprotocol}
An $(n,M_1,M_2)$ CSK generation protocol $\calC$, which specifies how two keys $(K_1,K_2)\in[M_1]\times[M_2]$ are generated with $n$ channel uses, consists of
\begin{itemize}
\item $n$ functions for Alice $\bg^{\rmA} = (g_1^{\rmA},\cdots, g_n^{\rmA})$, where $g_i^{\rmA}:\calF^{2i-2} \times \calR_{\rmA} \to \calX_1$ specifies how Alice chooses an channel input $X_{1,i}$ at time $i\in[n]$ using local randomness and previous public messages;
\item $n$ functions for Bob $\bg^{\rmB} = (g_1^{\rmB},\cdots, g_n^{\rmB})$, where $g_i^{\rmB}:\calF^{2i-2} \times \calR_{\rmB} \to \calX_2$ specifies how Bob chooses an channel input $X_{2,i}$ at time $i\in[n]$ using local randomness and previous public messages;
\item $n$ encoding functions for Alice $\mathbf{f}^{\rmA} = (f_1^{\rmA},\cdots, f_n^{\rmA})$, where $f_i^{\rmA}:\calF^{2i-2} \times \calX_1^i \times \calR_{\rmA} \to \calF$ specifies how Alice chooses a public message $F_{2i-1}$ at time $i\in[n]$ using previous public messages, previous channel inputs and local randomness;
\item $n$ encoding functions for Bob $\mathbf{f}^{\rmB} = (f_1^{\rmB},\cdots, f_n^{\rmB})$, where $f_i^{\rmB}:\calF^{2i-2} \times \calX_2^i \times \calR_{\rmB} \to \calF$ specifies how Bob chooses a public message $F_{2i}$ at time $i\in[n]$ using previous public messages, previous channel inputs and local randomness;
\item a key extraction function $\phi^{\rmA}:\calX_1^n \times\calF^{2n} \times \calR_{\rmA} \to [M_1]$ for Alice, which specifies how Alice generates the key $K_1$ using channel inputs $\bX_1$, all public messages and local randomness;
\item a key extraction function $\phi^{\rmB}:\calX_2^n \times\calF^{2n} \times \calR_{\rmB} \to [M_2]$ for Bob, which specifies how Bob generates the key $K_2$ using channel inputs $\bX_2$, all public messages and local randomness;
\item a key extraction function $\phi^{\rmc}:\calY^n \times \calF^{2n} \times \calR_{\rmC} \to [M_1]\times[M_2]$ for Charlie, which specifies how Charlie generates the keys $(K_1,K_2)$ using channel outputs $\bY$, all public messages and local randomness.
\end{itemize}
\end{definition}
}

{Using an $(n,M_1,M_2)$ CSK generation protocol $\calC$, the CSK generation process is specified as follows in an sequential manner. Fix any $i\in[n]$. Alice generates a channel input $X_{1,i}$ using local randomness $R_\rmA$, previous public messages $F^{2i-2}$ and the function $g_i^{\rmA}(F^{2i-2},R_\rmA)$. Analogously, Bob generates a channel input $X_{2,i}$ using $g_i^{\rmB}(F^{2i-2},R_\rmB)$. Charlie obtains a channel output $Y_i$ via the MAC $W_{Y|X_1X_2}$ and Willie obtains $Z_i$ via the MAC $W_{Z|X_1X_2}$. To reduce ambiguity between $(X_{1,i},X_{2,i})$ and $Y_i$, Alice transmits a public message $F_{2i-1}$ using previous public messages $F^{2i-2}$, channel inputs till now $X_1^i$ and local randomness via the function $f_i^{\rmA}(F^{2i-2},X_1^i,R_\rmA)$ while Bob transmits $F_{2i}=f_i^{\rmB}(F^{2i-2},X_2^i,R_\rmB)$.}

{Let $\bF$ denote all public messages $\bF:=(F_1,\cdots,F_{2n})$. The channel inputs of Alice and Bob are $\bX_1=(X_{1,1},\ldots,X_{1,n})$ and $\bX_2=(X_{2,1},\ldots,X_{2,n})$, respectively. The observed sequences of Charlie and Willie are $\bY=(Y_1,\ldots,Y_n)$ and $\bZ=(Z_1,\ldots,Z_n)$, respectively. Using $(\bF,\bX_1,R_\rmA)$, Alice generates the secret key $K_1$ via $\phi^{\rmA}(\bF,\bX_1,R_\rmA)$ while Bob generates the secret key $K_2$ via $\phi^{\rmB}(\bF,\bX_2,R_\rmB)$. Using $(\bF,\bY,R_\rmC)$, Charlie generates keys $(\hatK_1,\hatK_2)$ via $\phi^{\rmC}(\bF,\bY,R_\rmC)$. Let the joint distribution of $(\bX_1,\bX_2,\bY,\bZ,K_1,K_2,\hatK_1,\hatK_2)$ be denoted by $\hatP_{\bX_1\bX_2\bY\bZ K_1K_2\hatK_1\hatK_2\bF}$, and let all other distributions $\hatP_{\cdot}$ be induced by this joint distribution.}

\begin{remark}
In our key generation protocol, Charlie is a passive receiver who transmits nothing. This is in stark contrast to the point-to-point case in \cite{Tahmasbi2017cns}, where the legitimate receiver transmits public messages. The reason why we choose a different direction of communication is to enable multiple key generation. Otherwise, technical challenges arise since one needs to study channel resolvability over a broadcast channel. This is because the reverse channel model of a MAC with two transmitters and one receiver is a broadcast channel with one transmitter and two receivers. The point-to-point case does not suffer this problem since the reverse channel model for a point-to-point channel is still a point-to-point channel.
\end{remark}

\begin{remark}\label{remark:oneway}
{
Under the covertness constraint, it is preferred to adopt non-interactive public discussion, as described in \cite{Tahmasbi2017cns}. To illustrate the need for such a model, consider the following application scenario for CSK generation. A submarine conducting a secret mission needs to generate secret keys covertly with its allies on shore. Since the submarine cannot disclose its existence or location, it cannot transmit any information. In this case, the submarine can act as Charlie.}
\end{remark}

\subsection{Performance Metric}\label{subsection:PerformanceMetric}

Fix four symbols $(x_{10},x_{11},x_{20},x_{21})\in\bbN_{+}^{4}$. For ease of analysis, we consider binary input alphabets: $\calX_1=\left\{x_{10},x_{11}\right\}$ and $\calX_2=\left\{x_{20},x_{21}\right\}$. We choose $x_{10}$ and $x_{20}$ as the innocent symbols, which are continuously transmitted by Alice and Bob, respectively, when no meaningful communication takes place. The assumption of binary input can be easily generalized to arbitrary finite input alphabet, following \cite[Section VII-B]{Bloch2016Resolv}. Given $y\in\calY$ and $z\in\calZ$, define the following probabilities:
\begin{align}
P_0(y)&:=W_{Y|X_1X_2}(y|x_{10},x_{20}),\label{eq:def_covertPro_P0}\\
Q_0(z)&:=W_{Z|X_1X_2}(z|x_{10},x_{20}).\label{eq:def_covertPro_Q0}
\end{align}
Analogously, $P_{1}$ and $Q_{1}$ are defined as the conditional distributions $W_{Y|X_1X_2}$ and $W_{Z|X_1X_2}$ with input $(x_{11},x_{20})$; $P_{2}$ and $Q_{2}$ are defined as conditional distributions with inputs  $(x_{10},x_{21})$; $P_3$ and $Q_3$ are defined as conditional distributions with inputs $(x_{11},x_{21})$.

Intuitively, $P_0$ and $Q_0$ correspond to the output distributions of sequences observed by Charlie and Willie, respectively, when no meaningful symbols are transmitted. 
Consistent with \cite[Section III]{Bloch2016Resolv}, we assume that i) for each $i\in[3]$, $P_i\ll P_0$, $Q_i\ll Q_0$, and ii) $Q_0$ cannot be represented as a linear combination of the other $Q_i$ for $i\in[3]$. Otherwise, the problem degenerates since CSK generation is either be impossible, or can be trivially achieved with a positive rate.

{
The performance of a CSK generation protocol is evaluated via reliability \eqref{eq:metric_reliability}, secrecy \eqref{eq:metric_secrecy} and covertness \eqref{eq:metric_covert}. Fix any positive real numbers $(\varepsilon,\delta,\tau)\in\bbR_+^{3}$. An $(n,M_1,M_2)$ CSK generation protocol $\calC$ per Definition \ref{def:KGprotocol} is called an $(n,M_1, M_2,\varepsilon,\delta,\tau)$ protocol if
\begin{align}
P_\rme(\calC)&:=\P{\hatK_1\neq K_1 \text{ or } \hatK_2\neq K_2}\leq\varepsilon,\label{eq:metric_reliability}\\
S(\calC)&:=\D{\hatP_{K_1K_2\bF \bZ}}{P^{\rmU}_{K_1}\times P^{\rmU}_{K_2}\times P^{\rmU}_{\bF}\times\hatP_{\bZ}}\leq\delta,\label{eq:metric_secrecy}\\
L(\calC)&:=\D{\hatP_{\bZ}}{Q_0^{n}}\leq\tau,\label{eq:metric_covert}
\end{align}
where $P^{\rmU}_{K_1}$, $P^{\rmU}_{K_2}$ and $P^{\rmU}_{\bF}$ are uniform distributions defined on $[M_1]$, $[M_2]$ and $\calF^{2n}$, respectively. The constraints \eqref{eq:metric_reliability} and \eqref{eq:metric_covert} are standard reliability and covertness constraints. The constraint \eqref{eq:metric_secrecy} represents a strong secrecy constraint that can be decomposed as follows:
\begin{align}
&\D{\hatP_{K_1K_2\bF \bZ}}{P^{\rmU}_{K_1}\times P^{\rmU}_{K_2}\times P^{\rmU}_{\bF}\times\hatP_{\bZ}}\nn\\*
&=\D{\hatP_{K_1K_2\bF \bZ}}{\hatP_{K_1}\times \hatP_{K_2}\times \hatP_{\bF}\times\hatP_{\bZ}}
+\D{\hatP_{K_1}}{P^{\rmU}_{K_1}}
+\D{\hatP_{K_2}}{P^{\rmU}_{K_2}}
+\D{\hatP_{\bF}}{P^{\rmU}_{\bF}},
\end{align}
requiring all keys and public messages to be uniformly distributed and independent of Willie's observation when $\delta$ is small enough.}

{
The capacity region of CSK generation is defined as follows.
\begin{definition}\label{def:csk}
A CSK rate pair $(R_1, R_2)\in\bbR_+^2$ is achievable if there exists a sequence of $\{(n,M_{{1n}}, M_{{2n}},\varepsilon_n,\delta_n,\tau_n)\}_{n\in\bbN_+}$ protocols such that
\begin{align}
&\lim_{n\to\infty}\varepsilon_n
=\lim_{n\to\infty}\delta_n
=\lim_{n\to\infty}\tau_n
=0,\\
&\log M_{{1n}}=\omega(\log n),\ 
\log M_{{2n}}=\omega(\log n),
\end{align}
and
\begin{align}
\liminf_{n\to\infty}\frac{\log M_{{1n}}}{\sqrt{n\tau_n}}\geq R_1,\ 
\liminf_{n\to\infty}\frac{\log M_{{2n}}}{\sqrt{n\tau_n}}\geq R_2.
\end{align}
The convex closure of the set of all achievable CSK rate pairs is called the CSK capacity region and denoted as $C_{\mathrm{csk}}$.
\end{definition}
}

{When the covertness constraint is removed in \eqref{eq:metric_covert}, our problem reduces to WSK generation over a MAC. Such a problem serves as intermediate results of our achievability analysis. For completeness, the capacity region of WSK generation is given as follows.
\begin{definition}\label{def:wsk}
A WSK rate pair $(R_1, R_2)\in\bbR^2_{+}$ is achievable if there exists a sequence of $\{(n,M_{{1n}}, M_{{2n}},\varepsilon_n, \delta_n)\}_{n\in\bbN_+}$ protocols such that
\begin{align}
&\lim_{n\to\infty}\varepsilon_n
=\lim_{n\to\infty}\delta_n
=0,
\end{align}
and
\begin{align}
\liminf_{n\to\infty}\frac{\log M_{{1n}}}{n}\geq R_1,\ 
\liminf_{n\to\infty}\frac{\log M_{{2n}}}{n}\geq R_2.
\end{align}
The convex closure of the set of all achievable WSK rate pairs is called the WSK capacity region and denoted as $C_{\mathrm{wsk}}$.
\end{definition}
}

\section{Result and Discussions}
\label{section:MainResult}

{
\subsection{Covert Secret Key Generation}\label{subsection:mainresult_CSKG}
Let $\calB:=\{0,1\}$. Given any distribution $\boldsymbol{\rho}\in\calP(\calB)$, define the following functions:
\begin{align}
\zeta(z)&:=\sum_{i\in[2]}\rho_i (Q_i(z)-Q_0(z)),~z\in\calZ,\label{eq:def_zeta}\\
\chi(\boldsymbol{\rho})&:=\sum_z \frac{\zeta^2(z)}{Q_0(z)},\label{eq:def_chi}\\
\kappa(\boldsymbol{\rho})&:=\sqrt{\frac{2}{\chi(\boldsymbol{\rho})}}\label{def:kappa}.
\end{align}
Furthermore, to facilitate the statement of the theorem, we define the following two sets:
\begin{align}
\label{eq:covert_region_1}
\calR_{\mathrm{in}}(\boldsymbol{\rho})
&:=\Big\{(R_1,R_2)\in\bbR_+^2:~\forall~i\in[2],
R_i\leq \rho_i\kappa(\boldsymbol{\rho})
\left\{\D{P_i}{P_0}-\D{Q_i}{Q_0}\right\}^{+}\Big\},\\
\label{eq:covert_region_2}
\calR_{\mathrm{out}}(\boldsymbol{\rho})
&:=\Big\{(R_1,R_2)\in\bbR_+^2:~\forall~i\in[2],
R_i \leq \rho_i
\kappa(\boldsymbol{\rho})\D{P_i}{P_0}\Big\}.
\end{align}

\begin{theorem}\label{theorem:main}
The capacity region $C_{\mathrm{csk}}$ for CSK generation satisfies
\begin{align}
\bigcup_{\boldsymbol{\rho}\in\calP(\calB)} \calR_{\mathrm{in}}(\boldsymbol{\rho})
\subseteq C_{\mathrm{csk}}\subseteq 
\bigcup_{\boldsymbol{\rho}\in\calP(\calB)} \calR_{\mathrm{out}}(\boldsymbol{\rho}).
\end{align}
\end{theorem}
The achievability and converse proofs of Theorem \ref{theorem:main} are provided in Sections \ref{section:mainproof_ach} and \ref{section:mainproof_con}, respectively.
}

We make the following remarks. Firstly, without the covertness constraint, the capacity region for the secret key generation problem usually involves more than three bounds when two keys are generated. For example, \cite[Theorem 2]{Huishuai2017Cellular} characterizes the capacity region for multiple private key generation problem with an untrusted helper. The region includes three bounds: two bounds constrain individual key rates $R_1$ and $R_2$, respectively, and one additional bound on the sum rate $R_1+R_2$. However, when the covertness constraint is imposed, Theorem~\ref{theorem:main} involves only two individual rate bounds, while the bound on the sum rate disappears. This phenomenon was discussed intuitively in \cite{Arumugam2019MAC}: ``because covertness is such a stringent constraint that the covert users never transmit \emph{enough bits} to saturate the capacity of the channel".

Secondly, Theorem \ref{theorem:main} generalizes the point-to-point setting in~\cite{Tahmasbi2017cns} to establish capacity region for multiterminal CSK generation protocols. In the achievability part, we propose an auxiliary problem, characterize the performance of the auxiliary problem and clarify its connection to the CSK generation problem. This way, the performance analysis of CSK generation can be decomposed into five parts: source simulation, reliability, secrecy, covertness, and key rate. The first three parts are analyzed via the theoretical techniques for channel coding and channel resolvability while the last two parts are analyzed by carefully designing the channel inputs to satisfy the covertness constraint. In the converse part, we modify the converse result in \cite{Csiszar2004Source,Csiszar2008Channel} by imposing the covertness constraint. In the original converse result without the covertness constraint, there are five bounds for generating two keys among three parties, two of which are satisfied automatically due to the relationship between marginal rate and the sum rate. When the covertness constraint is imposed, one of the remaining three bounds becomes inactive.

{Thirdly, although Theorem \ref{theorem:main} holds for three legitimate parties, the results in Theorem \ref{theorem:main} can be generalized to arbitrary finite number of legitimate parties. The achievability proof can be generalized by appropriately choosing the user set concerning the auxiliary problem in Lemma \ref{lemma:auxi}. The converse proof can be done similarly to the current proof, but the single-letterization step can be complicated since the number of bounds increases exponentially with respect to the number of keys to be generated.  
}

{Finally, the capacity regions of CSK generation and covert communication over the same MAC is dual to each other, as discussed in detail in Section \ref{subsection:duality}. In a nutshell, the rates of the secret keys generated in our problem and the rates of secret keys required for covert communication both correspond to the rate gaps to achieve reliability \cite[Appendix D]{Arumugam2019MAC} and resolvability \cite[Appendix E]{Arumugam2019MAC}, respectively.}

\begin{table}[tb]
\centering
\caption{Numerical calculation parameters}
\label{tab:parameters}
\begin{tabular}{ccccc}
\hline
$(x_1,x_2)$&$(0,0)$&$(1,0)$&$(0,1)$&$(1,1)$\\
\hline
$W^{(1)}_{Y|X_1X_2}(1|x_1,x_2)$& 0.67 & 0.10 & 0.27 & 0.56\\
$W^{(1)}_{Z|X_1X_2}(1|x_1,x_2)$& 0.33 & 0.62 & 0.48 & 0.15\\
\hline
$W^{(2)}_{Y|X_1X_2}(1|x_1,x_2)$& 0.1 & 0.3 & 0.2 & 0.9\\
$W^{(2)}_{Z|X_1X_2}(1|x_1,x_2)$& 0.3 & 0.4 & 0.4 & 0.8\\
\hline
\end{tabular}
\end{table}

{
We provide the following two numerical examples to illustrate our results. Let $x_{10}=x_{20}=0$ and $x_{11}=x_{21}=1$.
\begin{example}
Consider two MACs $W^{(1)}_{Y|X_1X_2}$ and $W^{(1)}_{Z|X_1X_2}$ with parameters in the first two lines of Table \ref{tab:parameters}. Note that $\D{P_{1}}{P_0}>\D{Q_{1}}{Q_0}$ and $\D{P_{2}}{P_0}>\D{Q_{2}}{Q_0}$. The inner and outer bounds for the covert capacity region $C_{\mathrm{csk}}$ in Theorem \ref{theorem:main} are plotted in Fig.~\ref{fig:capacity}, together with the rate region for  a specific choice of $\boldsymbol{\rho}=\boldsymbol{\rho^*}=(\rho_1^*,\rho_2^*)=(0.28,0.72)$. 
\end{example}

\begin{figure}[tb]
\centering
\includegraphics[width=0.5\columnwidth]{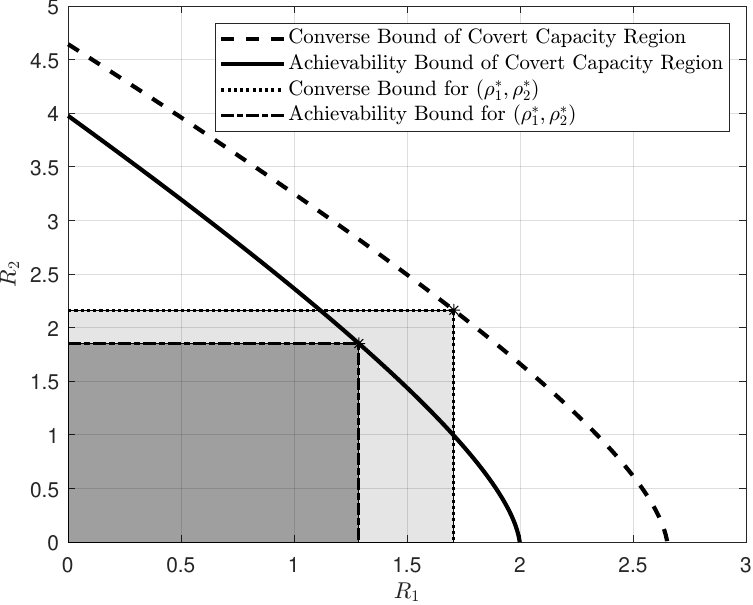}
\caption{Plot of inner and outer bounds for CSK capacity region $C_{\mathrm{csk}}$ over MACs with channel matrices $W^{(1)}_{Y|X_1X_2}$ and $W^{(1)}_{Z|X_1X_2}$ specified in Table \ref{tab:parameters}.}
\label{fig:capacity}
\end{figure}

\begin{example}
The second example concerns symmetric MACs $W^{(2)}_{Y|X_1X_2}$ and $W^{(2)}_{Z|X_1X_2}$, whose parameters are given in the last two lines of Table \ref{tab:parameters}. In this case,  $Q_i(z)-Q_0(z)$ are equal for each $i\in[2]$, and $\chi(\boldsymbol{\rho})=0.0476$, which is independent of $\boldsymbol{\rho}$. As a result, the inner bound can be achieved by time-division. However, time division is not necessarily optimal since the inner and outer bounds do not match. Similar discussion for covert communication over a MAC can be found in \cite[Remark 1]{Arumugam2016MAC}. 
\end{example}
}

\subsection{Wiretap Secret Key Generation}\label{subsection:wiretapKG}
\begin{figure}[tb]
\centering
\includegraphics[width=0.6\columnwidth]{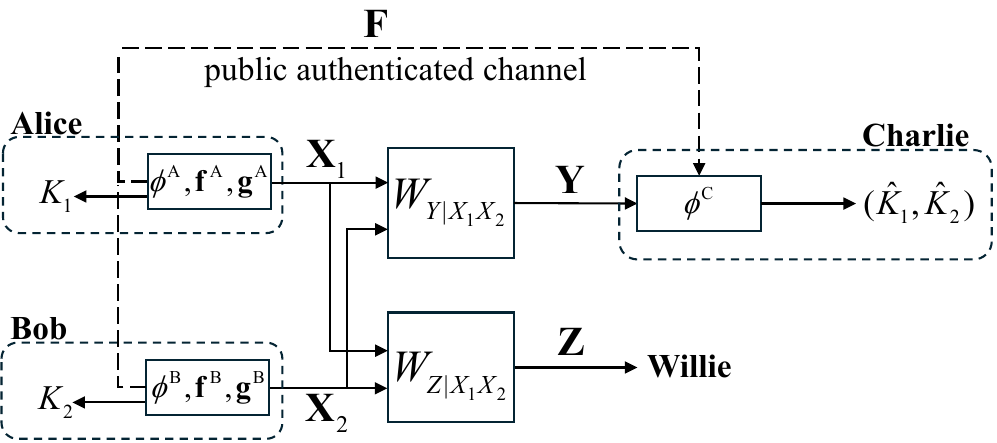}
\caption{System model for wiretap secret key generation over multiple access channels with non-interactive public discussion.}
\label{fig:wiretap_model}
\end{figure}

{When the covertness constraint is removed, CSK generation specializes to WSK generation as shown in Fig. \ref{fig:wiretap_model}. When there is only one transmitter, i.e., when Bob is absent and $X_2$ is a constant, the capacity $C_{\mathrm{wsk}}$ was bounded~\cite{Ahlswede1993KG, Maurer1993KG} as follows:
\begin{align}
\sup_{P_X}\{I(X;Y)-I(X;Z)\}\leq C_{\mathrm{wsk}}\leq \sup_{P_X}I(X;Y|Z)\label{eq:wiretap_p2p}
\end{align}
}

{
As an intermediate step to prove the capacity region for CSK generation, we bound the capacity region of WSK generation over the MAC, which generalizes the point-to-point result with one secret key in \eqref{eq:wiretap_p2p} to the multiterminal case in Fig. \ref{fig:wiretap_model} where two secret keys are generated. Fix $\calU=[2]$. Recall that $X_\calU=\{X_i\}_{i\in\calU}$. Let $P_{X_\calU}\in\calP(\calX^{|\calU|})$ be an arbitrary distribution. To present our result, define the following two sets:
\begin{align}
\calR^{\prime}_{\mathrm{in}}(P_{X_\calU})
:=\bigg\{R_{\calU}&:\forall~\emptyset\neq\calT\subseteq\calU,
\sum_{i\in\calT}R_{i}\leq 
I(X_{\calT};Y|X_{\calT^{c}})-I(X_{\calT};Z)
\bigg\},\label{eq:wiretap_1}\\
\calR^{\prime}_{\mathrm{out}}(P_{X_\calU})
:=\Big\{R_{\calU}&:\forall~\emptyset\neq\calT\subseteq\calU,
\sum_{i\in\calT}R_{i}\leq
I(X_{\calT};Y,X_{\calT^{c}}|Z)
\Big\}\label{eq:wiretap_2},
\end{align}
where the mutual information terms are calculated with respect to the distributions induced by $P_{X_{\calU}}$ and the MACs $W_{Y|X_1X_2}$ and $W_{Z|X_1X_2}$.}

\begin{corollary}\label{corollary:wiretap}
The capacity region $C_{\mathrm{wsk}}$ for WSK generation satisfies
\begin{align}
\sup_{P_{X_\calU}\in\calP(\calX^{|\calU|})}\calR^{\prime}_{\mathrm{in}}(P_{X_\calU})
\subseteq C_{\mathrm{wsk}}\subseteq
\sup_{P_{X_\calU}\in\calP(\calX^{|\calU|})}\calR^{\prime}_{\mathrm{out}}(P_{X_\calU}).
\end{align}
\end{corollary}
The proof sketch of Corollary \ref{corollary:wiretap} is provided in Appendix \ref{appendix:wiretap}.

{We make the following remarks. Firstly, the inner bound $\calR^{\prime}_{\mathrm{in}}(P_{X_\calU})$ can be seen as the rate gaps of the results of (i) channel coding over a two-user MAC \cite[Section 4.5]{ElGamal2011Network}, corresponding to conditions $\sum_{i\in\calT}{R_{i}}\leq I(X_{\calT};Y|X_{\calT^{c}})$ and (ii) channel resolvability over a two-user MAC \cite[Remark 1]{Helal2020Resolv}, corresponding to conditions $\sum_{i\in\calT}{R_{i}}\geq I(X_{\calT};Z)$, for any nonempty set $\calT\subseteq\calU$.}

{Secondly, Corollary \ref{corollary:wiretap} holds for any finite number $k$ of legitimate users by setting $\calU=[k]$. Our results complement \cite{Gohari2023Upper} by providing an inner bound to WSK capacity region. When specialized to the point-to-point setting, the capacity bound in \eqref{eq:wiretap_p2p} can be recovered from Corollary \ref{corollary:wiretap} by setting $\calU=[1]$, $X_1=X$ and $X_2$ equals a constant.}

{Finally, we discuss the difference between the results of CSK and WSK generation. In a nutshell, it boils down to the impact of the covertness constraint. For CSK, covert communication plays an important role while channel coding is critical for WSK generation. Recall that the covert communication differs from the standard channel coding problem in that the number of symbols transmitted over $n$ channel users scales in the order of $O(\sqrt{n})$, which is much less than $O(n)$ for channel coding without the covertness constraint. This explains the differences in capacity regions for CSK and WSK generation. In CSK generation, to deal with the difficulty introduced by low-weight channel inputs enforced by the covertness constraint, the proof of Theorem \ref{theorem:main} requires combining non-asymptotic bounds of channel reliability and resolvability \cite{Arumugam2019MAC} with \emph{stronger} inequalities \cite{Arumugam2019MAC} (cf. Lemma 5), such as Bernstein's inequality in Lemma \ref{lemma:Bernstein}. In contrast, bounds for WSK capacity region in Corollary \ref{corollary:wiretap} can be established combining results of channel reliability \cite{ElGamal2011Network} and resolvability \cite{Helal2020Resolv} with simpler concentration inequalities, such as Hoeffding's inequality.
}

\section{Achievability Proof of Theorem \ref{theorem:main}}
\label{section:mainproof_ach}
Traditional achievability proofs, e.g.,~\cite{Bloch2011PhysicalLayer,Maurer2000PA}, employ information reconciliation and privacy amplification to generate secret keys, relying on concentration inequalities for conditional entropies. However, as noted in \cite{Tahmasbi2017cns}, achieving covertness necessitates an alternative approach that uses concentration inequalities for mutual information. To address this gap, the authors of \cite{Tahmasbi2017cns} introduced an auxiliary problem and utilized a likelihood encoder. However, in the point-to-point setting \cite{Tahmasbi2017cns}, the terminals controlling the channel input is separate from the one transmitting public information, complicating extensions to the multiterminal case. In our proof, we design a key generation protocol where the same terminals perform both tasks, thereby generalizing the point-to-point framework to a multiterminal setting.

To establish the achievability part of Theorem~\ref{theorem:main}, we construct a key generation protocol that satisfies the reliability, secrecy, and covertness constraints in \eqref{eq:metric_reliability}-\eqref{eq:metric_covert}. Firstly, to ensure covertness, we define and analyze the properties of a covert process. Secondly, to analyze the reliability and secrecy constraints, we introduce an auxiliary problem that achieves reliability and resolvability and leverages the auxiliary problem to design a key generation protocol using likelihood encoders for the original problem. Finally, we discuss the duality of the theoretical benchmarks between covert key generation and covert communication over a two-user MAC.

\subsection{Covert Process}\label{subsection:properties}
In this section, we introduce the covertness process in \cite[Section IV]{Arumugam2019MAC}, which specifies a sequence of probability distributions that helps achieve covertness. Fix any positive sequence $\{\alpha_n\}_{n\in\bbN_{+}}\in o\big(\frac{1}{\sqrt{n}}\big) \cap \omega\big(\frac{\log n}{n}\big)$. An example of $\alpha_n$ is $\{\frac{1}{\log n\sqrt{n}}\}_{n\in\bbN_+}$.

\begin{definition}[Covert Process]\label{def:covertProcess}
{Fix any $i\in[2]$ and positive real numbers $(\rho_1,\rho_2)\in(0,1)^2$ such that $\rho_1+\rho_2=1$. Define the input distribution such that}
\begin{align}
Q_{X_i}(x_{i1})=1-Q_{X_i}(x_{i0})=\rho_i \alpha_n.\label{eq:def_covertPro_QX}
\end{align}
The output distributions $(Q_{Y},Q_{Z})$ induced by input distributions $Q_{X_1}$ and $Q_{X_2}$ and two MACs $W_{Y|X_1X_2}$ and $W_{Z|X_1X_2}$ satisfy that for $(y,z)\in\calY\times\calZ$, 
\begin{align}
Q_{Y}(y)&:=\sum_{(x_1,x_2)\in\calX_1\times\calX_2} 
W_{Y|X_1X_2}(y|x_1,x_2)\prod_{i\in[2]}Q_{X_i}(x_i),\label{eq:def_covertPro_QY}\\
Q_{Z}(z)&:=\sum_{(x_1,x_2)\in\calX_1\times\calX_2} 
W_{Z|X_1X_2}(z|x_1,x_2)\prod_{i\in[2]}Q_{X_i}(x_i).\label{eq:def_covertPro_QZ}
\end{align}
The corresponding product distributions are:
\begin{align}
Q^n_{X_i}=\prod_{j=1}^{n}Q_{X_i},\ 
Q^n_{Y}=\prod_{j=1}^{n}Q_{Y},\ 
Q^n_{Z}=\prod_{j=1}^{n}Q_{Z}.
\end{align}
\end{definition}

\begin{table}[tb]
\centering
\scriptsize
\caption{commonly used notations}
\label{tab:notations}
\begin{tabular}{cc}
\hline
Notation & Corresponding Definition\\
\hline
$\alpha_n$ & A designed sequence with value in $(0,1)$\\
$x_{10}$ & The innocent symbol in the input alphabet of $Q_{X_1}$ \\
$x_{11}$ & The meaningful symbol in the input alphabet of $Q_{X_1}$ \\
$x_{20}$ & The innocent symbol in the input alphabet of $Q_{X_2}$ \\
$x_{21}$ & The meaningful symbol in the input alphabet of $Q_{X_2}$ \\
$Q_{X_1}$ & Channel input distribution such that $Q_{X_1}(x_{11})=\rho_1 \alpha_n$ \\
$Q_{X_2}$ & Channel input distribution such that $Q_{X_2}(x_{21})=\rho_2 \alpha_n$ \\
$W_{Y|X_1X_2}$ & Conditional distribution of MAC $W_{Y|X_1X_2}\in\calP(\calY|\calX_1\calX_2)$ \\
$W_{Z|X_1X_2}$ & Conditional distribution of MAC $W_{Z|X_1X_2}\in\calP(\calZ|\calX_1\calX_2)$\\
$Q_Y$ & Channel output distribution defined by $W_{Y|X_1X_2}$ and $Q_{X_1}Q_{X_2}$ \\
$Q_Z$ & Channel output distribution defined by $W_{Z|X_1X_2}$ and $Q_{X_1}Q_{X_2}$ \\
$P_0$ & Channel output distribution $W_{Y|X_1=x_{10},X_2=x_{20}}$ \\
$P_1$ & Channel output distribution $W_{Y|X_1=x_{11},X_2=x_{20}}$ \\
$P_2$ & Channel output distribution $W_{Y|X_1=x_{10},X_2=x_{21}}$ \\
$P_3$ & Channel output distribution $W_{Y|X_1=x_{11},X_2=x_{21}}$ \\
$Q_0$ & Channel output distribution $W_{Z|X_1=x_{10},X_2=x_{20}}$ \\
$Q_1$ & Channel output distribution $W_{Z|X_1=x_{11},X_2=x_{20}}$ \\
$Q_2$ & Channel output distribution $W_{Z|X_1=x_{10},X_2=x_{21}}$ \\
$Q_3$ & Channel output distribution $W_{Z|X_1=x_{11},X_2=x_{21}}$ \\
\hline
\end{tabular}
\end{table}

For convenience, Table \ref{tab:notations} summarizes the commonly used notations defined above. The covert process can be viewed as low-weight, independent, and identically distributed (i.i.d.) input distributions transmitted through noisy multiple-access channels. As shown in Lemma \ref{lemma:nEstimate}, the resulting output distribution $Q^n_{Z}$ remains asymptotically indistinguishable from the innocent distribution $Q^n_{0}$.

Consistent with \cite[Section IV]{Arumugam2019MAC}, we assume that $\frac{Q_{X_1}(x_{11})}{Q_{X_2}(x_{21})}=\frac{\rho_1}{\rho_2}$ in \eqref{eq:def_covertPro_QX}. Fix $\boldsymbol{\rho}=(\rho_1,\rho_2)$, the vector $\boldsymbol{\rho}$ quantifies the relative weighting of the codewords generated by $Q_{X_1}$ and $Q_{X_2}$. This setting is introduced to precisely quantify the fraction of channel uses in which legitimate parties transmit meaningful symbols $x_{i1}$ for each $i\in[2]$, without introducing the specific key generating protocols.
For $z\in\calZ$, {analogously to \eqref{eq:def_zeta}-\eqref{eq:def_chi}, we also define the following two functions:}
\begin{align}
\zeta_n(z)&:=
\frac{Q_{Z}(z)-Q_0(z)}{\alpha_n},\label{eq:def_zeta_n}\\
\chi_n(\boldsymbol{\rho})&:=\sum_z \frac{\zeta_n^2(z)}{Q_0(z)}.\label{eq:def_chi_n}
\end{align}

{In the following lemma, the first- and second-order Taylor expansions are presented for mutual information terms between the MAC inputs and outputs, as a function of  $\alpha_n$. Fix any non-empty subset $\calT\subseteq\calU=[2]$. Note that all mutual information terms in this section are defined for the RVs $(X_{\calU},Y,Z)\in\calX^{|\calU|}\times\calY\times\calZ$, whose joint distribution is specified in Definition \ref{def:covertProcess}.}

\begin{lemma}\label{lemma:nEstimate}
For $n\in\bbN_{+}$ large enough, the following hold:

\begin{enumerate}[i)]
\item \label{eq:lemma_nEstimate_1}
Let $Q_Z$ and $Q_0$ be defined as per \eqref{eq:def_covertPro_QZ} and \eqref{eq:def_covertPro_Q0}, respectively. It follows that
\begin{align}
\D{Q_Z}{Q_0}\label{eq:I_a1}
&=\frac{1}{2}\alpha_n^2\chi_n(\boldsymbol{\rho})+O(\alpha_n^3).
\end{align}

\item \label{eq:lemma_nEstimate_2}
For all $z \in \calZ$, $\lim_{n\to\infty} \zeta_n(z) = \zeta(z)$ and $\lim_{n\to\infty} \chi_n(\boldsymbol{\rho}) = \chi(\boldsymbol{\rho})$. Furthermore,
\begin{align}
I\left(X_{\calT};Y|X_{\calT^{\rmc}}\right) &= \sum_{t \in \calT} \rho_t \alpha_n \D{P_{t}}{P_0} +O(\alpha_n^2),\\
I\left(X_{\calT};Z\right) &= \sum_{t \in \calT} \rho_t \alpha_n \D{Q_{t}}{Q_0} +O(\alpha_n^2)\label{eq:lemma_mutual_z},
\end{align}
and
\begin{align}
\Var{\log\frac{W_{Y|X_{\calU}}}{W_{Y|X_{\calT^{\rmc}}}}}&=O(\alpha_n),\\
\Var{\log\frac{W_{Z|X_{\calT}}}{Q_{Z}}}&=O(\alpha_n).
\end{align}

\item \label{eq:lemma_nEstimate_3}
{
There exists a positive constant $C$ independent of $n$ such that 
a) if for some $x_{\calU}\in\calX^{|\calU|}$ and $y\in\calY$, $Q_{X_{\calU}Y}(x_{\calU},y)>0$, 
\begin{align}
\left|
\log\frac{W_{Y|X_{\calU}}(y|x_{\calU})}{W_{Y|X_{\calT^{\rmc}}}(y|x_{\calT^{\rmc}})}
-I\left(X_{\calT};Y|X_{\calT^{\rmc}}\right)
\right|&\leq C,
\end{align}
and b) if for some $x_{\calT}\in\calX^{|\calT|}$ and $z\in\calZ$, $Q_{X_{\calT}Z}(x_{\calT},z)>0$,
\begin{align}
\left|
\log\frac{W_{Z|X_{\calT}}(z|x_{\calT})}{Q_{Z}(z)}
-I\left(X_{\calT};Z\right)
\right|&\leq C.
\end{align}
}

\item \label{eq:lemma_nEstimate_4}
Furthermore, if the Markov chain $Y-X_\calU-Z$ holds,
\begin{align}
I\left(X_{\calT};Y,Z|X_{\calT^{\rmc}}\right)=
\sum_{t\in\calT}\rho_t \alpha_n 
\big(\D{P_{t}}{P_0}&+\D{Q_{t}}{Q_0}\big) +O(\alpha_n^2).
\end{align}
\end{enumerate}
\end{lemma}
{
The results in \ref{eq:lemma_nEstimate_1})-\ref{eq:lemma_nEstimate_3}) were proved in \cite[Lemma 1, Appendices D and E]{Arumugam2019MAC}, and the result in \ref{eq:lemma_nEstimate_4}) was proved in Appendix~\ref{appendix:proof_lemma_iv:mutualInfo}. Using Lemma \ref{lemma:nEstimate}, we can analyze the covertness constraint and bound the key rates, as shown in Section \ref{subsection:KGScheme} and Appendix \ref{appendix:proof_lemma:auxi}, respectively.
}

\subsection{Auxiliary Coding Scheme}\label{subsection:auxiliary}
\begin{figure}[tb]
\centering
\includegraphics[width=0.6\columnwidth]{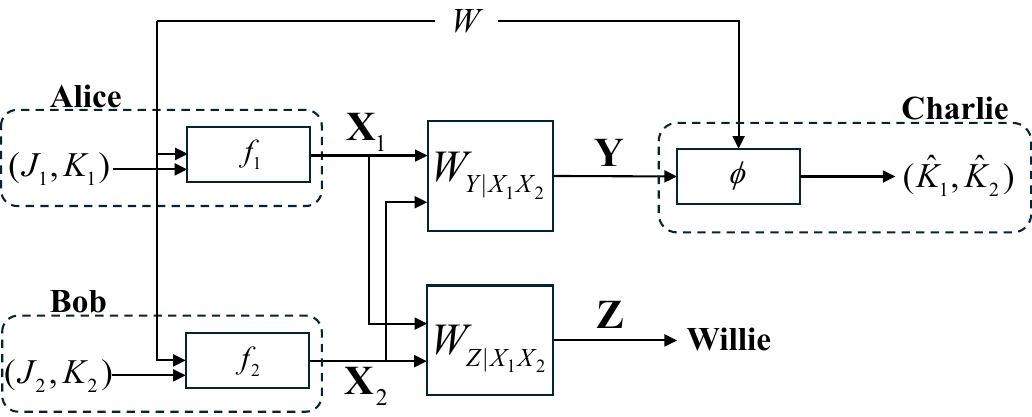}
\caption{\label{fig:auxi_model}System model for the auxiliary coding problem.}
\end{figure}

The key generation problem can be effectively addressed through an auxiliary MAC coding framework. This approach allows us to leverage established results from channel coding and resolvability. Figure \ref{fig:auxi_model} illustrates the system model for this auxiliary coding problem. The primary distinction between the auxiliary coding model and the key generation framework lies in the processing order of the secret keys. In the auxiliary coding scheme, legitimate parties first generate the keys locally and then encode them into codewords. Conversely, in the key generation protocol, codewords are generated according to input distributions, and keys are subsequently decoded from these sequences.

In this auxiliary problem, three types of messages are processed. Intuitively, $W$ presents the public message used for information reconciliation between the legitimate users, $J$ provides additional randomness to simulate the output distribution of Willie's observation when no key is generated, $K$ represents the secret key to be established.

Let $({G},M_1,M_2,N_{1},N_{2})\in\bbN_{+}^{5}$. The communication process begins when Charlie generates a message $W$ uniformly distributed over $[{G}]$ and transmits it over a public noiseless channel. Alice possesses two messages: $J_1$ and $K_1$, uniformly distributed over $[N_{1}]$ and $[M_1]$ respectively. Similarly, Bob has messages $J_2$ and $K_2$, uniformly distributed over $[N_{2}]$ and $[M_2]$ respectively. The coding scheme consists of:
\begin{itemize}
\item Alice's Encoder 1: $f_1:[{G}]\times[M_1]\times[N_{1}]\rightarrow \calX_1^n$;
\item Bob's Encoder 2: $f_2:[{G}]\times[M_2]\times[N_{2}]\rightarrow \calX_2^n$;
\item Charlie's Decoder: $\phi: [{G}]\times\calY^n\rightarrow[M_1]\times[M_2]$.
\end{itemize}

Alice encodes her three messages into the codeword $\bX_1=f_1(W, J_1, K_1)$, while Bob encodes his into $\bX_2=f_2(W, J_2, K_2)$. These codewords are transmitted through the MACs $W_{Y|X_1X_2}$ and $W_{Z|X_1X_2}$. Finally, Charlie receives $\bY$ and decodes $(\hatK_1, \hatK_2)=\phi(W, \bY)$. 
Let $\tilde{P}_{\bX_1\bX_2\bY\bZ K_1K_2\hatK_1\hatK_2J_1J_2W}$ denote the joint distribution induced by the code $(f_1,f_2,\phi)$, where
\begin{align}
&\tilde{P}_{\bX_1\bX_2\bY\bZ K_1K_2\hatK_1\hatK_2J_1J_2W}\nn\\*
&=P^{\rmU}_{K_1}\times P^{\rmU}_{K_2}\times P^{\rmU}_{J_1}\times P^{\rmU}_{J_2}\times P^{\rmU}_{W}\times\tilde{P}_{\bX_1|K_1J_1W}\label{eq:def_tildeP}
\times\tilde{P}_{\bX_2|K_2J_2W}\times W^n_{Y|X_1X_2}\times W^n_{Z|X_1X_2}\times\tilde{P}_{\hatK_1\hatK_2|W\bY},
\end{align}
where all distributions $\tilde{P}$ are induced by this joint distribution. Note that $\tilde{P}_{W}=P^{\rmU}_{W}$ is uniformly distributed over $[{G}]$, as are $\tilde{P}_{K_1},\tilde{P}_{K_2},\tilde{P}_{J_1},\tilde{P}_{J_2}$.

The following lemma presents an achievability result for this auxiliary problem. Let $W=(W_1,W_2)$, we have

\begin{lemma}\label{lemma:auxi}
Let $(n,G,M_1,M_2,N_{1},N_{2})\in\bbN_{+}^{6}$. For positive real numbers $(\mu_1,\mu_2,\mu_3)\in\bbR_{+}^3$, nonempty set $\calT\subseteq\calU=[2]$ and each $i\in[2]$, if we set
\begin{align}
\log N_{{\calT}}+\log M_{{\calT}}
&=(1-\mu_1)nI(X_{\calT};Y|X_{\calT^{\rmc}}),\label{eq:exp_cond1}\\
\log N_{{\calT}}
&=(1+\mu_2)nI(X_{\calT};Z),\label{eq:exp_cond2}\\
\log {G}_{i}+\log N_{i}+\log M_{i}
&=(1+\mu_3)nH(X_i),\label{eq:exp_cond3}
\end{align}
there exists a sequence of codes $\left\{\left(f_{1n}, f_{2n}, \phi_n\right)\right\}_{n\geq 1}$ and a positive constant $\xi\in\bbR_{+}$ such that
\begin{align}
\label{eq:exp_1}
\P[\tilde{P}]{\hatK_1\neq K_1 \text{ or } \hatK_2\neq K_2}
&\leq
\exp\left(-\xi n\alpha_n\right),\\
\label{eq:exp_2}
\V{\tilde{P}_{K_1K_2W\bZ}}
{\tilde{P}_{K_1}
\times \tilde{P}_{K_2}
\times \tilde{P}_{W}
\times Q_{Z}^{n}}
&\leq
\exp\left(-\xi n\alpha_n\right),\\
\label{eq:exp_3}
\V{\tilde{P}_{\bX_1}}{Q_{X_1}^{n}}
&\leq
\exp\left(-\xi n\alpha_n\right),\\
\label{eq:exp_4}
\V{\tilde{P}_{\bX_2}}{Q_{X_2}^{n}}
&\leq
\exp\left(-\xi n\alpha_n\right).
\end{align}
\end{lemma}
This lemma establishes four essential properties: reliability \eqref{eq:exp_1}, secrecy \eqref{eq:exp_2}, and distribution similarity \eqref{eq:exp_3}-\eqref{eq:exp_4}. The latter two ensure that the distribution induced by our coding scheme closely approximates the distribution induced by the key generation protocol. The proof is given in Appendix \ref{appendix:proof_lemma:auxi}.

\subsection{Key Generation Scheme}\label{subsection:KGScheme}
In this subsection, we construct a key generation protocol that satisfies Definition \ref{def:KGprotocol} using the auxiliary coding scheme developed previously. 
Three conditional probabilities obtained from the auxiliary scheme induced distribution $\tilde{P}$ in \eqref{eq:def_tildeP} are used in the proof here: $\tilde{P}_{W_1K_1J_1|\bX_1}$ used as Alice's encoding function as well as key extractor; $\tilde{P}_{W_2K_2J_2|\bX_2}$ used as Bob's encoding function as well as key extractor; $\tilde{P}_{\hatK_1\hatK_2|\bY W}$ used as Charlie's key extractor.

The key generation protocol $\calC$ is defined as following. First Alice generates random sequence $\bX_1$ according to $Q_{X_1}^{n}$, similarly Bob generates $\bX_2$ according to $Q_{X_2}^{n}$. They then send them through the MACs $W_{Y|X_1X_2}$ and $W_{Z|X_1X_2}$. After the $n^{th}$ transmission, Alice uses the sequence $\bX_1$ and $\tilde{P}_{W_1K_1J_1|\bX_1}$ to sample $W_1,K_1,J_1$, and then transmit $W_1$ through the public channel. Similarly, Bob uses $\bX_2$ and $\tilde{P}_{W_2K_2J_2|\bX_2}$ to sample $W_2,K_2,J_2$ and transmit $W_2$. Finally, Charlie receives sequence $\bY$ and use the likelihood encoder $\tilde{P}_{\hatK_1\hatK_2|\bY W}$ to recover $(\hatK_1,\hatK_2)$. Recall that the distribution induced by the key generation protocol $\calC$ be $\hatP_{\bX_1\bX_2\bY\bZ K_1K_2\hatK_1\hatK_2J_1J_2W}$, with all marginal distributions $\hatP_{\cdot}$ derived from this joint distribution. Note that all the stochastic coding functions for the key generation protocol in Definition \ref{def:KGprotocol} are induced by this joint distribution.

We divide the following analysis into five parts: source simulation, reliability, secrecy, covertness, and throughput analysis. 
The source simulation part bounds the distance between the distributions induced by the auxiliary coding scheme and the key generation protocol. This allows us to analyze the throughput with the help of Lemma \ref{lemma:auxi}. The remaining three parts of reliability, secrecy and covertness correspond to \eqref{eq:metric_reliability}-\eqref{eq:metric_covert}.

Firstly, we analyze the source simulation part, begin by expressing the distribution $\hatP$ induced by the key generation protocol:
\begin{align}
&\hatP_{\bX_1\bX_2\bY\bZ K_1K_2\hatK_1\hatK_2J_1J_2W}\nn\\*
&=Q_{X_1}^{n}\times
Q_{X_2}^{n}\times
W_{Y|X_1X_2}^{n}\times
W_{Z|X_1X_2}^{n}
\times
\tilde{P}_{W_1K_1J_1|\bX_1}\times
\tilde{P}_{W_2K_2J_2|\bX_2}\times
\tilde{P}_{\hatK_1\hatK_2|W\bY}\label{eq:ssa_2}\\
&=Q_{X_1}^{n}\times
Q_{X_2}^{n}\times
\tilde{P}_{\bY\bZ K_1K_2\hatK_1\hatK_2J_1J_2W|\bX_1\bX_2}\label{eq:ssa_2_ad},
\end{align}
where \eqref{eq:ssa_2} follows from the likelihood encoders induced by the auxiliary problem, and \eqref{eq:ssa_2_ad} follows from the definition of $\tilP$ in \eqref{eq:def_tildeP}.
Similarly, $\tilde{P}$ can be written as
\begin{align}
&\tilde{P}_{\bX_1\bX_2\bY\bZ K_1K_2\hatK_1\hatK_2J_1J_2W}\nn\\*
&=\tilde{P}_{\bX_1}\times\tilde{P}_{\bX_2}\times
\tilde{P}_{\bY\bZ K_1K_2\hatK_1\hatK_2J_1J_2W|\bX_1\bX_2}\label{eq:ssa_3}.
\end{align}

Denoted as $R(\hatP,\tilde{P})$, the difference between the two induced distribution can be bounded as
\begin{align}
R(\hatP,\tilde{P})
&:=\V{\hatP_{\bX_1\bX_2\bY\bZ K_1K_2\hatK_1\hatK_2J_1J_2W}}
{\tilde{P}_{\bX_1\bX_2\bY\bZ K_1K_2\hatK_1\hatK_2J_1J_2W}}\nn\\
&=\V{Q_{X_1}^{n}\times Q_{X_2}^{n}}
{\tilde{P}_{\bX_1}\times\tilde{P}_{\bX_2}}\label{eq:ssa_22}\\
&=\V{Q_{X_1}^{n}}{\tilde{P}_{\bX_1}}+
\V{Q_{X_2}^{n}}{\tilde{P}_{\bX_2}}\\
&\leq\exp\left(-\xi n\alpha_n\right)\label{eq:ssa_24},
\end{align}
where \eqref{eq:ssa_22} follows from \eqref{eq:ssa_2_ad} and \eqref{eq:ssa_3}, which implies that the TV distance between the two induced distribution comes from the input distribution. The source simulation term \eqref{eq:ssa_24} follows from the results in \eqref{eq:exp_3}, \eqref{eq:exp_4} and the symmetry of TV distance in its parameters $\V{P}{Q}=\V{Q}{P}$. 
This analysis shows that the difference between the distribution induced by our key generation protocol and that of the auxiliary coding scheme is bounded by $\exp(-\xi n\alpha_n)$, which is vanishingly small for large $n$.

Then the reliability, secrecy and covertness parts are analyzed. 
Follow from the definition of reliability metric in \eqref{eq:metric_reliability}, we bound the error probability as
\begin{align}
P_\rme(\calC)
&=\P[\hatP]{\hatK_1\neq K_1 \text{ or } \hatK_2\neq K_2}\\
&=\P[\tilde{P}]{\hatK_1\neq K_1 \text{ or } \hatK_2\neq K_2}+R(\hatP,\tilde{P})\\
&\leq\exp\left(-\xi n\alpha_n\right)+\exp\left(-\xi n\alpha_n\right)\label{eq:ra_1}\\
&\leq\exp\left(-\xi^{\prime} n\alpha_n\right),
\end{align}
where the the first term of \eqref{eq:ra_1} using the results of the auxiliary problem $\tilde{P}$ from \eqref{eq:exp_1}, while the second term arises from the difference between the auxiliary and original problems, as given in \eqref{eq:ssa_24}. 

Follows from the definition of secrecy metric in \eqref{eq:metric_secrecy}, there exist a constant $\xi^{\prime\prime}>0$ such that
\begin{align}
&\V{\hatP_{\bZ K_1K_2W}}
{\hatP_{\bZ}\times P_{W}^{\rmU}\times P_{K_1}^{\rmU}\times P_{K_2}^{\rmU}}\nn\\*
&\leq\V{\hatP_{\bZ K_1K_2W}}{\tilde{P}_{\bZ K_1K_2W}}
+\V{\tilde{P}_{\bZ K_1K_2W}}
{\hatP_{\bZ}\times P_{W}^{\rmU}\times P_{K_1}^{\rmU}\times P_{K_2}^{\rmU}}\label{eq:sa_2}\\
&\leq\exp\left(-\xi n\alpha_n\right)+\exp\left(-\xi n\alpha_n\right)\label{eq:sa_3}\\
&\leq\exp\left(-\xi^{\prime\prime} n\alpha_n\right)\label{eq:sa_4},
\end{align}
where \eqref{eq:sa_2} comes from the triangle inequality of the TV distance, and $\hatP_{\bZ}=Q_{Z}^{n}$. The first term in \eqref{eq:sa_3} comes from \eqref{eq:ssa_24} and the second term comes from \eqref{eq:exp_2}. 
Then there exist a constant $\xi^{\prime\prime}>0$ such that
\begin{align}
S(\calC)
&=\D{\hatP_{\bZ K_1K_2W}}
{\hatP_{\bZ}\times P_{W}^{\rmU}\times P_{K_1}^{\rmU}\times P_{K_2}^{\rmU}}\label{eq:sa_5}\\
&\leq
\V{\hatP_{\bZ K_1K_2W}}
{\hatP_{\bZ}\times P_{W}^{\rmU}\times P_{K_1}^{\rmU}\times P_{K_2}^{\rmU}}
\log\left(M_1M_2G\right)
+H_{b}\left(2\V{\hatP_{\bZ K_1K_2W}}
{\hatP_{\bZ}\times P_{W}^{\rmU}\times P_{K_1}^{\rmU}\times P_{K_2}^{\rmU}}
\right)\label{eq:approx_constraint_ineq}
\\&\leq
\V{\hatP_{\bZ K_1K_2W}}
{\hatP_{\bZ}\times P_{W}^{\rmU}\times P_{K_1}^{\rmU}\times P_{K_2}^{\rmU}}
\left(O(n\alpha_n)+\log\frac{e}{2\V{\hatP_{\bZ K_1K_2W}}
{\hatP_{\bZ}\times P_{W}^{\rmU}\times P_{K_1}^{\rmU}\times P_{K_2}^{\rmU}}}
\right)\label{eq:approx_constraint_log}\\
&\leq\exp\left(-\xi^{\prime\prime} n\alpha_n\right)\label{eq:sa_6},
\end{align}
where \eqref{eq:approx_constraint_ineq} follows from 
\begin{align}
\D{P_{XY}}{P^{\rmU}_{X}\times P_{Y}}
&\leq\V{P_{XY}}{P^{\rmU}_{X}\times P_{Y}}
\log\left|\calX\right|
+H_{b}\left(2\V{P_{XY}}{P^{\rmU}_{X}\times P_{Y}}\right),
\end{align}
in \cite[Problem 17.1]{Csiszarbook}, and \eqref{eq:approx_constraint_log} follows from \eqref{eq:exp_cond3} and \eqref{eq:def_covertPro_QX} that $\log\left(M_1M_2G\right)$ scales as $O(n\alpha_n)$, and follows \cite[Equation (69)]{Tahmasbi2020KGActive} by $H_{b}(x)\leq x\log\frac{e}{x}$. Hence, when $n$ is large enough, we have \eqref{eq:sa_6} from \eqref{eq:sa_4}, which is vanishing. 

Follow from the definition of covertness metric in \eqref{eq:metric_covert}, the covertness term is bounded as
\begin{align}
L(\calC)
&=\D{\hatP_{\bZ}}{Q_0^{n}}\\
&=\D{Q_{Z}^{n}}{Q_0^{n}}\\
&=\frac{n}{2}\alpha_n^2\chi_n(\boldsymbol{\rho})+O(n\alpha_n^3),\label{eq:ca_2}
\end{align}
where \eqref{eq:ca_2} comes from Claim \ref{eq:lemma_nEstimate_1}) of Lemma \ref{lemma:nEstimate}.

Finally, we analyze the achievable key rates. For the individual key rates, we obtain
\begin{align}
\frac{\log M_{{1}}}{\sqrt{nL(\calC)}}
&\geq
\frac{n\left\{(1-\mu_1)I(X_1;Y|X_2)-(1+\mu_2)I(X_1;Z)\right\}^{+}}{n\sqrt{
\frac{1}{2}\alpha_n^2\chi_n(\boldsymbol{\rho})+O(\alpha_n^3)
}}\label{eq:ta_2}\\
&=\rho_1\sqrt{\frac{2}{\chi_n(\boldsymbol{\rho})}}
\frac{\left\{\D{P_{1}}{P_0}-\D{Q_{1}}{Q_0}\right\}^{+}}
{\sqrt{1+O(\alpha_n)}}\label{eq:ta_3}\\
&=\rho_1\kappa(\boldsymbol{\rho})
\left\{\D{P_{1}}{P_0}-\D{Q_{1}}{Q_0}\right\}^{+},\label{eq:ta_4}
\end{align}
where \eqref{eq:ta_2} follows from \eqref{eq:exp_cond1} and \eqref{eq:exp_cond2} of the auxiliary scheme in Lemma \ref{lemma:auxi}, \eqref{eq:ta_3} follows from Claim \ref{eq:lemma_nEstimate_2}) of Lemma \ref{lemma:nEstimate}, \eqref{eq:ta_4} follows from $\lim_{n\to\infty}\chi_n(\boldsymbol{\rho})=\chi(\boldsymbol{\rho})$, and $O(\alpha_n)$ vanishes when $n$ is large enough. 
Similarly, $\frac{\log M_{{2}}}{\sqrt{nL(\calC)}}$ can be bounded as follows:
\begin{align}
&\frac{\log M_{2}}{\sqrt{nL(\calC)}}
\geq\rho_2\kappa(\boldsymbol{\rho})
\left\{\D{P_{2}}{P_0}-\D{Q_{2}}{Q_0}\right\}^{+}.\label{eq:ta_7}
\end{align}
For the sum term, we have
\begin{align}
\frac{\log M_{{1}}+\log M_{{2}}}{\sqrt{nL(\calC)}}
&\geq
\frac{n\left\{(1-\mu_1)I(X_1,X_2;Y)-(1+\mu_2)I(X_1,X_2;Z)\right\}^{+}}{n\sqrt{
\frac{1}{2}\alpha_n^2\chi_n(\boldsymbol{\rho})+O(\alpha_n^3)
}}\label{eq:ta_8}\\
&=\sum_{i\in[2]}\rho_i\kappa(\boldsymbol{\rho})
\left\{\D{P_{i}}{P_0}-\D{Q_{i}}{Q_0}\right\}^{+},\label{eq:ta_9}
\end{align}
where \eqref{eq:ta_8} follows from Lemma \ref{lemma:auxi}, \eqref{eq:ta_9} follows from Claim \ref{eq:lemma_nEstimate_2}) of Lemma \ref{lemma:nEstimate}, and $n$ is large enough. Note that the sum rate constraint \eqref{eq:ta_9} is automatically satisfied given the separate rate constraints \eqref{eq:ta_4} and \eqref{eq:ta_7}, and becomes inactive when $n$ is large enough.

\subsection{Duality with Covert Communication}\label{subsection:duality}
We now examine the relationship between covert secret key (CSK) generation and covert communication, revealing an important duality between these problems.

In the covert communication problem, ``keyless" communication is only possible when $\D{P_i}{P_0}>\D{Q_i}{Q_0}$ for any $i\in\calU=[2]$. This condition indicates that the legitimate channel is \emph{better} then the wiretap channel. Otherwise, a shared key is required, with key rate characterized in \cite[Eq.(19)]{Arumugam2019MAC} as
\begin{align}
\Big\{
\{R_i\}_{i\in\calU}&:\forall i\in\calU,
R_i \geq \rho_i
\kappa(\boldsymbol{\rho})
\left\{\D{P_i}{P_0}-\D{Q_i}{Q_0}\right\}^{+}
\Big\}.
\end{align}

While in the CSK generation problem, we have shown in \eqref{eq:covert_region_1} that positive key rates are achievable only when $\D{P_i}{P_0}>\D{Q_i}{Q_0}$ for any $i\in\calU=[2]$:
\begin{align}
\Big\{
\{R_i\}_{i\in\calU}&:\forall i\in\calU,
R_i \leq \rho_i
\kappa(\boldsymbol{\rho})
\left\{\D{P_i}{P_0}-\D{Q_i}{Q_0}\right\}^{+}
\Big\}.
\end{align}

This highlights a fundamental duality: the expression $\rho_i\kappa(\boldsymbol{\rho})\left\{\D{P_i}{P_0}-\D{Q_i}{Q_0}\right\}^{+}$ represents both (i) the required key rate for covert communication and (ii) the achievable key generation rate for CSK. The term $\D{P_i}{P_0}-\D{Q_i}{Q_0}$ quantifies the information advantage of the legitimate channel over the wiretap channel.

This duality emerges from our analysis in Section \ref{subsection:auxiliary}, where we demonstrated that the achievable key rates are determined by the difference between reliability \eqref{eq:exp_cond1} and covertness requirements \eqref{eq:exp_cond2}. Specifically, our auxiliary coding scheme in Lemma \ref{lemma:auxi} shows that the key rate is constrained by $\log M_{\calT} \leq (1-\mu_1)nI(X_{\calT};Y|X_{\calT^{\rmc}}) - (1+\mu_2)nI(X_{\calT};Z)$, which directly leads to the rate expressions above when normalized by $\sqrt{nL(\calC)}$ and optimized. Similarly, in the analysis of covert communication~\cite[Eq. (27) and (29)]{Arumugam2019MAC} explains this gap.

\section{Converse Proof of Theorem \ref{theorem:main}}\label{section:mainproof_con}

We now derive the converse part of Theorem \ref{theorem:main}, building on the key capacity results over source and channel models by Csisz\'ar and Narayan \cite{Csiszar2004Source,Csiszar2008Channel}. 

First, we assume a fixed input distribution, allowing us to apply the same reasoning as in key generation over the multiterminal source model discussed in \cite{Csiszar2008Channel}. Notably, the converse for a PK generation model serves as an upper bound for the WSK generation model \cite[Theorem 4]{Csiszar2004Source}. This follows directly from the definition, as the WSK model can be viewed as a special case of the PK model.
Then for the sum term, we can treat Alice and Bob as a single super terminal, reducing the model to the PK generation problem analyzed in \cite[Theorem 2]{Csiszar2004Source}. Consequently, following \cite[Eq.(15)]{Csiszar2004Source}, the key size satisfies $\log M_{{1n}}+\log M_{{2n}}\leq \sum_{i=1}^{n}I(X_{i1},X_{i2};Y_i|Z_i)$. If only $K_1$ is generated, it is constrained to
\begin{align}
\log M_{{1n}}\leq \min\bigg\{&\sum_{i=1}^{n}I(X_{i1};Y_i,X_{i2}|Z_i),\sum_{i=1}^{n}I(X_{i1},X_{i2};Y_i|Z_i)\bigg\},
\end{align}
which is {simplified} $\log M_{{1n}}\leq \sum_{i=1}^{n}I(X_{i1};Y_i,X_{i2}|Z_i)$ due to the sum constraint. Similarly, when only $K_2$ is generated, we have $\log M_{{2n}}\leq \sum_{i=1}^{n}I(X_{i2};Y_i,X_{i1}|Z_i)$. As a result, let $\left\{\calC_n\right\}_{n\geq 1}$ be a sequence of $(n,M_{1n}, M_{2n}, \varepsilon_n, \delta_n, \tau_n)$ protocols, we obtain
\begin{align}
\log M_{{1n}}&\leq\sum_{i=1}^{n}I(X_{1i};Y_{i},X_{2i}|Z_i),\\
\log M_{{2n}}&\leq\sum_{i=1}^{n}I(X_{2i};Y_{i},X_{1i}|Z_i),\\
\log M_{{1n}}+\log M_{{2n}}&\leq \sum_{i=1}^{n}I(X_{1i},X_{2i};Y_{i}|Z_i).
\end{align}
Define RVs $(\bar{X_1},\bar{X_2},\bar{Y},\bar{Z})$, whose joint distribution depends on the average inputs $\bar{P}_{\rho_1\mu_{n}}:=\frac{1}{n}\sum_{i=1}^{n}\hatP_{X_{1i}}=\mathrm{Bern}(\rho_1\mu_{n})$ and $\bar{P}_{\rho_2\mu_{n}}:=\frac{1}{n}\sum_{i=1}^{n}\hatP_{X_{2i}}=\mathrm{Bern}(\rho_2\mu_{n})$, along with the conditional distribution $W_{YZ|X_1X_2}=W_{Y|X_1X_2}\times W_{Z|X_1X_2}$. 
We then bound $\log M_{{1}}$ as
\begin{align}
\nn&\sum_{i=1}^{n}I(X_{1i};X_{2i},Y_{i}|Z_i)\\*
&\leq nI(\barX_1;\barX_2,\barY|\barZ)\label{eq:conv_2}\\
&=n\left(I(\barX_1\barZ;\barX_2,\barY)-I(\barZ;\barX_2,\barY)\right)\\
&=n\left(I(\barZ;\barX_2,\barY|\barX_1)+I(\barX_1;\barX_2,\barY)-\left(I(\barY;\barZ|\barX_2)+I(\barX_2;\barZ)\right)\right)\\
&=n\left(\left(I(\barZ;\barY|\barX_1\barX_2)+I(\barZ;\barX_2|\barX_1)+I(\barX_1;\barY|\barX_2)+I(\barX_1;\barX_2)\right)-I(\barX_2;\barZ)-I(\barY;\barZ|\barX_2)\right)\label{eq:pc_3}\\
&=n\left(I(\barZ;\barX_2|\barX_1)+I(\barX_1;\barY|\barX_2)-I(\barX_2;\barZ)\right)\nn\\*
&\quad-n\left(I(\barX_1;\barY|\barX_2)+I(\barX_1;\barZ|\barX_2)-I(\barX_1;\barY\barZ|\barX_2)+I(\barY;\barZ|\barX_1\barX_2)\right)\label{eq:pc_4}\\
&=n\left(I(\barX_1;\barY,\barZ|\barX_2)-I(\barX_1;\barZ|\barX_2)+I(\barZ;\barX_2|\barX_1)-I(\barX_2;\barZ)\right)\label{eq:conv_3}\\
&=n\mu_n\left(\rho_1\left(\D{P_{1}}{P_0}+\D{Q_{1}}{Q_0}-\D{Q_{1}}{Q_0}\right)+\rho_2\left(\D{Q_{2}}{Q_0}-\D{Q_{2}}{Q_0}\right)\right)+O(n\mu_n^2)\label{eq:conv_4}\\
&=n\rho_1\mu_n\D{P_{1}}{P_0}+O(n\mu_n^2),\label{eq:conv_5}
\end{align}
where \eqref{eq:conv_2} follows from the concavity of $I(X_1,Y;X_2|Z)$ in $P_{X_1}$ and $P_{X_2}$, \eqref{eq:conv_4} follows from Claim \ref{eq:lemma_nEstimate_2}) and \ref{eq:lemma_nEstimate_4}) of Lemma \ref{lemma:nEstimate}, \eqref{eq:pc_4} and \eqref{eq:conv_3} follow since  $I(X_1;X_2)=0$ and $I(Y;Z|X_1,X_2)=0$ due to the independence of channel inputs and independence of two MAC channels, respectively.

The bound for $\log M_{{2}}$ is derived similarly using Lemma \ref{lemma:nEstimate}. The sum term follows as 
\begin{align}
&\sum_{i=1}^{n}I(X_{1i},X_{2i};Y_{i}|Z_i)\nn\\*
&\leq nI(\barX_1,\barX_2;\barY|\barZ)\label{eq:conv_2_1}\\
&=n\left(I(\barX_1,\barX_2;\barY,\barZ)-I(\barX_1,\barX_2;\barZ)\right)\label{eq:conv_2_2}\\
&=n\mu_n\bigg(\sum_{t\in[2]}
\rho_t\big(\D{P_{t}}{P_0}+\D{Q_{t}}{Q_0}-\D{Q_{t}}{Q_0}\big)\bigg)
+O(n\mu_n^2)\label{eq:conv_2_3}\\
&=n\mu_n\sum_{t\in[2]}\rho_t\D{P_{t}}{P_0}+O(n\mu_n^2)\label{eq:conv_2_4}.
\end{align}
Notably, \eqref{eq:conv_2_4} holds automatically for sufficiently large $n$ if the individual bounds on $\log M_{{1}}$ and $\log M_{{2}}$ are satisfied. Similarly to \cite[Eq.(60) and Eq.(61)]{Tahmasbi2017cns}, the proof is completed by following the same argument in \cite[Section V-C]{Arumugam2019MAC}.

\section{Conclusion}
\label{section:conclusion}
We established bounds on the capacity region of CSK generation over a two-user MAC with binary input, generalizing the results in \cite{Tahmasbi2017cns,Tahmasbi2020KGActive} from the point-to-point case with one secret key to the multiterminal case with multiple secret keys. Our results demonstrated the duality between CSK generation and covert communication over the same MAC, which corresponds to rate gaps between capacity regions for reliability and resolvability. Furthermore, when the covertness constraint is removed, we obtained bounds for the capacity region of multiterminal WSK generation over the same MAC and analyzed the impact of the covertness constraint. To obtain the theoretical theoretical results, we judiciously adapted channel resolvability and channel reliability results over the MAC in \cite{Arumugam2019MAC} and applied the converse technique in \cite{Csiszar2004Source,Csiszar2008Channel} with an additional covertness constraint.

There are several avenues for future research directions. Firstly, our bounds on the capacity region of CSK generation are not tight. It is valuable to tighten our bounds or find special cases of channels where the bounds match for a multiterminal setting. Secondly, we studied CSK generation over discrete MAC with finite input and output alphabets. In practice, the channel input and noise can both be continuous. Thus, it is worthwhile to generalize our results to continuous MAC. To do so, the techniques in \cite{Narayan2012GausConv,Zhou2020} can be helpful. 
Finally, regarding multiterminal CSK generation, while we focus on the MAC channel in this paper, it would also be interesting to investigate another important multiuser channel—the broadcast channel (BC)—and derive corresponding bounds on the CSK capacity region for this setting.
It was shown~\cite[Theorem 2]{Tan2019CovertBC} that time-division is optimal for covert communication over some BCs. A natural question is whether this result extends to CSK generation over a BC.

\appendix
\subsection{Proof of Claim \ref{eq:lemma_nEstimate_4}) of Lemma \ref{lemma:nEstimate}}\label{appendix:proof_lemma_iv:mutualInfo}

Recall the $\alpha_n$ and $\boldsymbol{\rho}=(\rho_1,\rho_2)$ defined in Section \ref{subsection:PerformanceMetric}. Let $Q_{YZ|X_1X_2}$ be denoted as $\Phi$, and define $\Phi_{0}$, $\Phi_{1}$ and $\Phi_{1}$ as the conditional distributions with inputs $(x_{10},x_{20})$, $(x_{11},x_{20})$ and $(x_{10},x_{21})$, respectively. For any nonempty set $\calT\subseteq\calU=[2]$, we have
\begin{align}
I\left(X_{\calT};Y,Z\right)
&=\sum_{t \in \calT} \rho_t \alpha_n \D{\Phi_{t}}{\Phi_{0}} +O(\alpha_n^2)\label{eq:app_c1_0}\\
&=\sum_{t \in \calT} \rho_t \alpha_n \D{P_{t}\times Q_{t}}{P_0\times Q_0} +O(\alpha_n^2)\label{eq:app_c1_1}\\
&=\sum_{t \in \calT} \rho_t \alpha_n \left(\D{P_{t}}{P_0}+\D{Q_{t}}{Q_0}\right) +O(\alpha_n^2),\label{eq:app_c1_2}
\end{align}
where \eqref{eq:app_c1_0} follows from replacing $Z$ with $(Y,Z)$ in \eqref{eq:lemma_mutual_z} in Lemma \ref{lemma:nEstimate}, \eqref{eq:app_c1_1} follows from $W_{YZ|X_1X_2}=W_{Y|X_1X_2}\times W_{Z|X_1X_2}$. Then the mutual information term has
\begin{align}
I\left(X_{\calT};Y,Z|X_{\calT^{\rmc}}\right)
&=I\left(X_{\calU};Y,Z\right)-I\left(X_{\calT^{\rmc}};Y,Z\right)\\
&=\sum_{t \in \calU} \rho_t \alpha_n \left(\D{P_{t}}{P_0}+\D{Q_{t}}{Q_0}\right)
-\sum_{t \in \calT^{\rmc}} \rho_t \alpha_n \left(\D{P_{t}}{P_0}+\D{Q_{t}}{Q_0}\right)
+O(\alpha_n^2)\label{eq:app_c1_3}\\
&=\sum_{t \in \calT} \rho_t \alpha_n \left(\D{P_{t}}{P_0}+\D{Q_{t}}{Q_0}\right) +O(\alpha_n^2),
\end{align}
where \eqref{eq:app_c1_3} follows from setting $\calT=\calU$ in \eqref{eq:app_c1_2}, and replacing $\calT$ with $\calT^{\rmc}$ in Claim \ref{eq:lemma_nEstimate_4}) of Lemma \ref{lemma:nEstimate}.

\subsection{Proof of Lemma \ref{lemma:auxi}}\label{appendix:proof_lemma:auxi}

Lemma \ref{lemma:auxi} establishes a coding scheme for the auxiliary coding problem in Fig. \ref{fig:auxi_model}, ensuring that the reliability, resolvability, and source simulation constraints in \eqref{eq:exp_1}-\eqref{eq:exp_4} are satisfied. These constraints are verified using a combination of non-asymptotic results and concentration inequalities.

\subsubsection{Reliability and Resolvability Proof}
Next lemma presents two non-asymptotic bounds on reliability and resolvability within a MAC model, adapted with slight modifications from the proofs in \cite[Appendices D and E]{Arumugam2019MAC}. Define the set $\calU:=[U]$ where $U\in\bbN_{+}$ and $U\geq 2$. Given a DM-MAC $W_{Y|X_{\calU}}\in\calP(\calY|\calX^{|\calU|})$ and encoders $f_i:[M_i]\rightarrow\calX_i^n$, while $M_{i}\in\bbN_{+}$ for each $i\in\calU$, we have

\begin{lemma}[Non-asymptotic Bounds]\label{lemma:nonAsymptotic}
We define $\hatP_{W_{\calU}X_{\calU}Y}$ as the distribution induced by messages $W_i$ uniformly distributed over $[M_i]$, respectively for each $i\in\calU$. Set $F=F_{\calU}$ as a set of random encoders such that $\left\{F_{i}(w_i)\right\}_{w_i\in[M_i]}$ are independently and uniformly distributed according to $Q_{X_i}$, and $\hatW_i$ is the optimal estimate of $W_i$ from $\bY$. The output distribution $Q_Y$ and the distribution induced by code $\hatQ_{\bY}$ are  
\begin{align}
Q_Y&:=
\sum_{\bX_{\calU}}
Q_{X_{\calU}}\left(\bx_{\calU}\right)
W_{Y|X_{\calU}}
\left(\by|\bx_{\calU}\right),\\
\hatQ_{\bY}&:=
\frac{1}{\prod_{i\in\calU}M_i}
\sum_{w_{\calU}}
W^{n}_{Y|X[W_{\calU}]}
\left(\by|\bx[w_{\calU}]\right),
\end{align}
Let $\mu>0$, define
\begin{align}
\gamma_{\calT}&:=(1-\mu)nI(X_{\calT};Y|X_{\calT^{\rmc}}),\label{eq:nonAsymptotic_define1}\\
\eta_{\calT}&:=(1+\mu)nI(X_{\calT};Y),\label{eq:nonAsymptotic_define2}
\end{align}
for any nonempty set $\calT\subseteq\calU$. Set $v_{\mathrm{min}}:=\min_{z\in\calZ}Q_0(z)$, we have
\begin{align}
\label{eq:nonAsymptotic_1}
\E[F]{\P{W_{\calU}\neq\hatW_{\calU}}}
&\leq\sum_{\substack{\calT\subseteq\calU:\\\calT\neq\emptyset}}
\Exp{-\gamma_{\calT}}\left(\prod_{i\in\calT}M_i\right)
+\sum_{\substack{\calT\subseteq\calU:\\\calT\neq\emptyset}}
\P{\sum_{i=1}^n
\log \frac{W_{Y|X_{\calU}}\left(Y|X_{\calU}\right)}
{W_{Y|X_{\calT^{\rmc}}}\left(Y|X_{\calT^{\rmc}}\right)}
<\gamma_{\calT}},\\
\label{eq:nonAsymptotic_2}
\E[F]{\D{\hatQ_{\bY}}{Q^{n}_{Y}}}
&\leq\sum_{\substack{\calT\subseteq\calU:\\\calT\neq\emptyset}}
\Exp{\eta_{\calT}}\frac{1}{\prod_{i\in\calT}M_i}\nn\\*
&\quad+n\log\left(\frac{2^U}{\prod_{u\in\calU}(1-\rho_u\alpha_u)v_{\min}}\right)\cdot
\sum_{\substack{\calT\subseteq\calU:\\\calT\neq\emptyset}}
\P{\sum_{i=1}^n\log \frac{W_{Y|X_{\calT}}\left(Y|X_{\calT}\right)}{Q_Y(Y)}
>\eta_{\calT}}.
\end{align}
\end{lemma}

In the model, there are two legitimate receivers, indexed by the set $\calU=[2]$ in Lemma \ref{lemma:nonAsymptotic}. For a fixed $n$, consider a set of random encoders $F=\left\{F_1,F_2\right\}$. The corresponding codewords $\left\{F\left(k_1,j_1,w\right)\right\}_{\left(k_1,j_1,w\right)\in[M_1]\times[N_{1}]\times[{G}]}$ and $\left\{F\left(k_2,j_2,w\right)\right\}_{\left(k_2,j_2,w\right)\in[M_2]\times[N_{2}]\times[{G}]}$ are drawn independently according to $Q_{X_1}$ and $Q_{X_2}$, respectively. 
If $\hatK_1$ is the optimal estimate of $K_1$, and $\hatK_2$ is the optimal estimate of $K_2$ from $\bY$ and $W$, then 
\begin{align}
&\E[F]{\P{\hatK_1\neq K_1 \text{ or } \hatK_2\neq K_2}}\nn\\*
&=\frac{1}{{G}}\sum_{w}
\E[F]{\P{\hatK_1\neq K_1 \text{ or } \hatK_2\neq K_2|W=w}}\\
&\leq
\sum_{\substack{\calT\subseteq[2]:\\\calT\neq\emptyset}}
\Exp{-\gamma_{\calT}}\left(\prod_{k\in\calT}M_{k}N_{k}\right)
+\sum_{\substack{\calT\subseteq[2]:\\\calT\neq\emptyset}}
\Pr\Bigg\{\sum_{i=1}^n
\log\frac{W_{Y|X_1X_2}\left(Y|X_1X_2\right)}
{W_{Y|X_{\calT^{\rmc}}}\left(Y|X_{\calT^{\rmc}}\right)}
<\gamma_{\calT}
\Bigg\}\label{eq:reliability_union},
\end{align}
Here, \eqref{eq:reliability_union} follows from Lemma \ref{lemma:nonAsymptotic}. We then introduce Bernstein's inequality to further refine the analysis.

\begin{lemma}[Bernstein's inequality]\label{lemma:Bernstein}
Let $\left\{U_i\right\}_{i=1}^n$ be independent zero-mean RVs such that $\abs{U_i} \leq c$ for a finite $c > 0$ almost surely for all $i\in[n]$. Then, for any $t>0$, 
\begin{align}
\P{\sum_{i=1}^n U_i > t} \leq \Exp 
{- \frac{\frac{1}{2}t^2}{\sum_{i=1}^n \E{U_i^2} + \frac{1}{3}ct}}. \label{eq:113}
\end{align}
\end{lemma}
Bernstein's inequality provides superior control over the tail probabilities under the covertness constraint than other concentration inequalities. In particular, Lemma \ref{lemma:Bernstein} allows us to establish vanishing bounds on the second term of \eqref{eq:reliability_union}, which is crucial for our asymptotic analysis.

For the second term in \eqref{eq:reliability_union}, following the definition of $\gamma_{\calT}$ in \eqref{eq:nonAsymptotic_define1}, when $\calT=\{2\}$, there exists a constant $\xi_1>0$ such that
\begin{align}
&\P{\sum_{i=1}^n
\log\frac{W_{Y|X_1X_2}\left(Y|X_1X_2\right)}{W_{Y|X_1}\left(Y|X_1\right)}
<(1+\mu)nI(X_2;Y|X_1)
}\nn\\*
&=\Pr\Bigg\{\sum_{i=1}^n\left(
\log\frac{W_{Y|X_1X_2}\left(Y|X_1X_2\right)}{W_{Y|X_1}\left(Y|X_1\right)}
-nI(X_2;Y|X_1)\right)
<\mu nI(X_2;Y|X_1)\Bigg\}\label{eq:reliability_1_0}\\
&\leq\Exp{-\frac{\frac{1}{2}\mu^2 n I(X_2;Y|X_1)^2}{
\var\big(\log\frac{W_{Y|X_1X_2}(Y|X_1X_2)}{W_{Y|X_2}(Y|X_2)}\big)
+\frac{C}{3}\mu I(X_2;Y|X_1)}}\label{eq:reliability_1_05}\\
&\leq\exp\left(-\xi_1 n\alpha_n\right),\label{eq:reliability_1_1}
\end{align}
where \eqref{eq:reliability_1_05} follows from Bernstein's inequality, \eqref{eq:reliability_1_1} follows from Claim \ref{eq:lemma_nEstimate_2}) of Lemma \ref{lemma:nEstimate}. The other two terms when $\calT=\{1\}$ and $\calT=\{1,2\}$ can be bounded similarly. Let $\mu_1>\mu>0$. For the first term in \eqref{eq:reliability_union}, we select appropriate values for $M_{{1}},M_{{2}},N_{{1}}$ and $N_{{2}}$ that satisfy
\begin{align}
\log N_{{1}}+\log M_{{1}}
&\leq(1-\mu_1)nI(X_1;Y|X_2),\label{eq:reliability_1_2_1}\\
\log N_{{2}}+\log M_{{2}}
&\leq(1-\mu_1)nI(X_2;Y|X_1),\\
\log N_{{1}}+\log M_{{1}}+\log N_{{2}}+\log M_{{2}}
&\leq(1-\mu_1)nI(X_1,X_2;Y),\label{eq:reliability_1_2_3}
\end{align}
then there exists a constant $\xi_2>0$ such that
\begin{align}
&\sum_{
\substack{\calT\subseteq[2]:\\\calT\neq\emptyset}}
\exp\left(-\gamma_{\calT}\right)
\left(\prod_{k\in\calT}M_{k}N_{k}\right)\nn\\*
&\leq\sum_{
\substack{\calT\subseteq[2]:\\\calT\neq\emptyset}}
\exp\big\{(\mu-\mu_1)nI(X_{\calT};Y|X_{\calT^{\rmc}})\big\}\label{eq:reliability_1_3}\\
&\leq\exp\left(-\xi_2 n\alpha_n\right),\label{eq:reliability_1_4}
\end{align}
where \eqref{eq:reliability_1_3} follows from the definition of $\gamma_{\calT}$ in \eqref{eq:nonAsymptotic_define1} and \eqref{eq:reliability_1_2_1}-\eqref{eq:reliability_1_2_3}, \eqref{eq:reliability_1_4} follows from Claim \ref{eq:lemma_nEstimate_2}) of Lemma \ref{lemma:nEstimate} and $\mu_1>\mu$. Thus, \eqref{eq:exp_1} is proved by combining \eqref{eq:reliability_1_1} and \eqref{eq:reliability_1_4}.

To prove the secrecy constraint in \eqref{eq:exp_2}, we have
\begin{align}   
&\E[F]{\D{\tilde{P}_{\bZ|K_1=k_1K_2=k_2W=w}}{Q_{Z}^{n}}}\nn\\*
&\leq\sum_{\substack{
\calT\subseteq[2]:\\
\calT\neq\emptyset
}}
\Exp{\eta_{\calT}}\frac{1}{\prod\limits_{k\in\calT}N_{k}}
+n\log\left(\frac{2^2}
{\prod\limits_{u\in\calU}(1-\rho_u\alpha_u)v_{\min}}
\right)\cdot\sum_{\substack{\calT\subseteq[2]:\\\calT\neq\emptyset}}
\P{\sum_{i=1}^n\log 
\frac{W_{Z|X_{\calT}}\left(Z|X_{\calT}\right)}{Q_Z(Z)}
>\eta_{\calT}
},\label{eq:secrecy_union}
\end{align}
where \eqref{eq:secrecy_union} follows from the second part of Lemma \ref{lemma:nonAsymptotic}. For the second term in \eqref{eq:secrecy_union}, following the definition of $\eta_{\calT}$ in \eqref{eq:nonAsymptotic_define2}, when $\calT=\{1\}$, there exists a constant $\xi_3>0$ such that
\begin{align}
&\P{\sum_{i=1}^n
\log \frac{
W_{Z|X_1}\left(Z|X_1\right)
}{Q_Z(Z)}
>(1+\mu)nI(X_1;Z)
}\nn\\*
&\leq\exp\left\{-\frac{\frac{1}{2}\mu^2 n I(X_1;Z)^2}
{\var\left(\log\frac{W_{Z|X_1}(Z|X_1)}{Q_{Z}(Z)}\right)
+\frac{C}{3}\mu I(X_1;Z)}
\right\}\label{eq:secrecy_step1_05}\\
&\leq\exp\left(-\xi_3 n\alpha_n\right),\label{eq:secrecy_step11}
\end{align}
where \eqref{eq:secrecy_step1_05} follows by invoking Bernstein's inequality, \eqref{eq:secrecy_step11} from Claim \ref{eq:lemma_nEstimate_2}) of Lemma \ref{lemma:nEstimate}. The other two terms when $\calT=\{2\}$ and $\calT=\{1,2\}$ can be bounded similarly.
Let $\mu_2>\mu>0$. For the first term in \eqref{eq:secrecy_union}, by choosing appropriate $N_{{1}}$ and $N_{{2}}$ that satisfy
\begin{align}
\log N_{{1}}
&\geq(1+\mu_2)nI(X_1;Z),\\
\log N_{{2}}
&\geq(1+\mu_2)nI(X_2;Z),\\
\log N_{{1}}+\log N_{{2}}
&\geq(1+\mu_2)nI(X_1,X_2;Z),
\end{align}
we obtain that there exists a constant $\xi_4>0$ such that
\begin{align}
\sum_{\substack{
\calT\subseteq[2]:\\
\calT\neq\emptyset
}}
\Exp{\eta_{\calT}}\frac{1}{\prod_{k\in\calT}N_{k}}
&\leq\exp\left(-\xi_4 n\alpha_n\right),\label{eq:secrecy_step1_2}
\end{align}
using a similar approach in \eqref{eq:reliability_1_3}. Combining \eqref{eq:secrecy_step11} and \eqref{eq:secrecy_step1_2} into \eqref{eq:secrecy_union}, we obtain that there exists a constant $\xi_5>0$ such that
\begin{align}
&\bbE_{F}\left[
\V{\tilde{P}_{K_1K_2W\bZ}}
{\tilde{P}_{K_1}
\times \tilde{P}_{K_2}
\times \tilde{P}_{W}
\times Q_{Z}^{n}}
\right]\nn\\*
&=\frac{1}{M_1M_2{G}}\sum_{k_1,k_2,w}\bbE_{F}\left[
\V{\tilde{P}_{\bZ|K_1=k_1K_2=k_2W=w}}
{Q_{Z}^{n}}
\right]\label{eq:secrecy_step22}\\
&\leq\exp\left(-\xi_5 n\alpha_n\right),\label{eq:secrecy_exp}
\end{align}
where\eqref{eq:secrecy_step22} follows from that $\tilde{P}_{K_1}$, $\tilde{P}_{K_2}$ and $\tilde{P}_{W}$ are uniformly distributed over their alphabets, \eqref{eq:secrecy_exp} follows from \eqref{eq:secrecy_union} and the Pinsker's inequality $\V{P}{Q}^2\leq \frac{1}{2}\D{P}{Q}$.
Thus, \eqref{eq:exp_2} is proved.

\subsubsection{Source Simulation Proof}
To prove \eqref{eq:exp_3} and \eqref{eq:exp_4}, we rely on the following two lemmas. Lemma \ref{lemma:oneShotResolvability} establishes a one-shot channel resolvability bound for a noiseless channel, derived directly from \cite[Lemma 1]{Tahmasbi2017cns}. Lemma \ref{lemma:Hoeffding} presents Hoeffding's inequality.

\begin{lemma}\label{lemma:oneShotResolvability}
Let $M\in\bbN_{+}$, given a message $W$ uniformly distributed over $[M]$ and an encoder $f:[M]\rightarrow\calX$, let $\hatP_X$ be the induced distribution $\hatP_X(x)=\frac{1}{M}\sum_{w}\bbI\left(f(w)=x\right)$. If $F$ is a random encoder such that $\left\{F(w)\right\}_{w\in[M]}$ are independent and identically distributed according to $P_X$, then for all $\gamma>0$, 
\begin{align}
\E[F]{\V{\hatP_X}{P_X}}
\leq\P{\log{\frac{1}{P_X(X)}}\geq\gamma}+\sqrt{\frac{\exp\left(\gamma\right)}{M}}.
\end{align}
\end{lemma}

\begin{lemma}[Hoeffding's inequality] \label{lemma:Hoeffding}
Let $\{U_i\}_{i=1}^n$ be a set of independent RVs such that $a_i \leq X_i \leq b_i$ almost surely, and let $U \triangleq \sum_{i=1}^n X_i$. For any $v > 0$,
\begin{align}
\P{|U - \E{U}| \geq v} \leq \exp\left(-\frac{2v^2}{\sum_{i=1}^n (b_i - a_i)^2}\right).
\end{align} 
\end{lemma}

Let $\mu_3>0$ and define $v_{1}:=\min_{x_1\in\calX_1}Q_{X_1}(x_1)$. Choose appropriate $G_{{1}}$, $G_{{2}}$, $M_{{1}}$, $M_{{2}}$, $N_{{1}}$, $N_{{2}}$ such that
\begin{align}
\log {G}_{1}+\log N_{{1}}+\log M_{{1}}=\left(1+\mu_3\right)nH(X_1),\\
\log {G}_{2}+\log N_{{2}}+\log M_{{2}}=\left(1+\mu_3\right)nH(X_2).
\end{align}
then we obtain that there exists a constant $\xi_6>0$ such that
\begin{align}
&\bbE_{F}\left[\V{\tilde{P}_{\bX_1}}{Q_{X_1}^{n}}\right]\nn\\*
&\leq\P{\sum_{i=1}^{n}\log{\frac{1}{Q_{X_1}(X_i)}}\geq\left(1+\frac{\mu_3}{2}\right)nH(X_1)}+\sqrt{\frac{\exp\left((1+\frac{\mu_3}{2})nH(X_1)\right)}{{G}_{1}N_{1}M_1}}\label{eq:sourceSimu_15}\\
&\leq\exp\left(-\frac{\mu_3^2 nH(X_1)}{2v_{1}^2}\right)
+\exp\left(-\frac{\mu_3}{2}nH(X_1)\right)\label{eq:sourceSimu_2}\\
&\leq\exp\left(-\xi_6 n\alpha_n\right),\label{eq:sourceSimu_3}
\end{align}
where \eqref{eq:sourceSimu_15} follows from Lemma \ref{lemma:oneShotResolvability}, \eqref{eq:sourceSimu_2} from Hoeffding's inequality, and \eqref{eq:sourceSimu_3} from $H(X_1)=(\rho_1\alpha_n)\log\frac{1}{\rho_1\alpha_n}+(1-\rho_1\alpha_n)\log\frac{1}{1-\rho_1\alpha_n}>\rho_1\alpha_n$ when $\alpha_n$ vanishes. The term $\bbE_{F}\left[\V{\tilde{P}_{\bX_2}}{Q_{X_2}^{n}}\right]$ can be bounded similarly. The source simulation part is simpler to the proof in \eqref{eq:reliability_union}, as it involves only the entropy term rather than the mutual information term.

\subsection{Proof of Corollary \ref{corollary:wiretap}}\label{appendix:wiretap}

By defining a similar auxiliary coding scheme, we first prove the following Lemma.
\begin{lemma}\label{lemma:auxi_wiretap}
Let $(n,G,M_1,M_2,N_{1},N_{2})\in\bbN_{+}^{6}$. For positive real numbers $(\mu_1,\mu_2,\mu_3)\in\bbR_{+}^3$, nonempty set $\calT\subseteq\calU=[2]$ and fixed distribution $Q_{X_i}$ for each $i\in[2]$, if we set
\begin{align}
\log N_{{\calT}}+\log M_{{\calT}}
&=(1-\mu_1)nI(X_{\calT};Y|X_{\calT^{\rmc}}),\label{eq:exp_cond1_wiretap}\\
\log N_{{\calT}}
&=(1+\mu_2)nI(X_{\calT};Z),\label{eq:exp_cond2_wiretap}\\
\log {G}_{i}+\log N_{i}+\log M_{i}
&=(1+\mu_3)nH(X_i),\label{eq:exp_cond3_wiretap}
\end{align}
there exists a sequence of codes $\left\{\left(f_{1n}, f_{2n}, \phi_n\right)\right\}_{n\geq 1}$ and a positive constant $\xi\in\bbR_{+}$ such that
\begin{align}
\label{eq:exp_1_wiretap}
\lim_{n\to\infty}\P[\tilde{P}]{\hatK_1\neq K_1 \text{ or } \hatK_2\neq K_2}=&0,\\
\label{eq:exp_2_wiretap}
\lim_{n\to\infty}\V{\tilde{P}_{K_1K_2W\bZ}}
{\tilde{P}_{K_1}
\times \tilde{P}_{K_2}
\times \tilde{P}_{W}
\times Q_{Z}^{n}}=&0,\\
\label{eq:exp_3_wiretap}
\lim_{n\to\infty}\V{\tilde{P}_{\bX_1}}{Q_{X_1}^{n}}=&0,\\
\label{eq:exp_4_wiretap}
\lim_{n\to\infty}\V{\tilde{P}_{\bX_2}}{Q_{X_2}^{n}}=&0.
\end{align}
\end{lemma}

The proof of Lemma \ref{lemma:auxi_wiretap} is much simpler than that of Lemma \ref{lemma:auxi}, as it does not require addressing low-weight codewords, and the mutual information terms are not of order $\alpha_n$, which vanishes asymptotically. Given \eqref{eq:exp_cond1_wiretap}, the constraint in \eqref{eq:exp_1_wiretap} is proved using the channel coding theorem for a MAC \cite[Section 4.5]{ElGamal2011Network}. Similarly, given \eqref{eq:exp_cond2_wiretap}, the constraint in \eqref{eq:exp_2_wiretap} is established using the theorem on resolvability for MAC with non-cooperating encoders \cite[Remark 1]{Helal2020Resolv}. The constraints \eqref{eq:exp_3_wiretap} and \eqref{eq:exp_4_wiretap} remain identical to those in Lemma \ref{lemma:auxi}. Based on Lemma \ref{lemma:auxi_wiretap}, Corollary \ref{corollary:wiretap} follows by applying the same steps as in Section \ref{section:mainproof_ach}.

\bibliographystyle{IEEEtran}
\bibliography{reference}
\end{document}